\documentclass[graybox, envcountchap]{svmult}

\usepackage{mathptmx}        % selects Times Roman as basic font
\usepackage{amsmath}
\usepackage{amssymb}
\usepackage{color}
\usepackage{helvet}          % selects Helvetica as sans-serif font
\usepackage{courier}         % selects Courier as typewriter font
\usepackage{dirtree}
\usepackage{booktabs}
\usepackage{makeidx}        % allows index generation
\usepackage{graphicx}        % standard LaTeX graphics tool
\usepackage{subfig}

\usepackage{multicol}        % used for the two-column index
\usepackage[bottom]{footmisc}% places footnotes at page bottom

\usepackage[misc]{ifsym}

\makeindex             % used for the subject index
\usepackage[%
  bookmarksopen=true,
  colorlinks=true,
  urlcolor=blue,
  linkcolor=blue,
  citecolor=blue
]{hyperref}

\begin{document}

%%%%%%%%%%%%%%%%%%%%%%%%%%%%%%%%%%%%%%%%%%%%%%%%%%%%%%%%%%%%%%%%%

\title{Black hole-neutron star binaries}
\author{Matthew D. Duez}
\institute{Matthew Duez (\Letter) \at Department of Physics \& Astronomy, Washington State University, Pullman, Washington 99164, USA, \email{m.duez@wsu.edu}}
%\and Second Author \at Name, Address of Institute \email{name@email.address}}
\maketitle

\abstract{
  The gravitational wave signals of black hole-neutron star (BHNS)
  binary systems have now been detected, and future detections might be
  accompanied by electromagnetic counterparts.  BHNS mergers involve much of
  the same physics as binary neutron star mergers:  strong gravity, nuclear
  density matter, neutrino radiation, and magnetic turbulence.  They also
  share with binary neutron star systems the potential for bright
  electromagnetic signals, especially gamma ray bursts and kilonovae, and
  the potential to be significant sources of r-process elements.  However,
  BHNS binaries are more asymmetric, and their mergers produce different
  amounts and arrangements of the various post-merger material components
  (e.g. disk and dynamical ejecta), together with a more massive
  black hole; these differences can have interesting consequences.  In this
  chapter, we review the modeling of BHNS mergers and post-merger evolution
  in numerical relativistic hydrodynamics and magnetohydrodynamics.  We
  attempt to give readers a broad understanding of the answers to the
  following questions.  What are the main considerations that determine
  the merger outcome?  What input physics must (or should) go into a BHNS
  simulation?  What have the most advanced simulations to date learned?
}

%%%%%%%%%%%%%%%%%%%%%%%%%%%%%%%%%%%%%%%%%%%%%%%%%%%%%%%%%

\section{Introduction}
\label{sec:intro}

Black hole-neutron star (BHNS) binary inspirals and mergers involve
extremes of all four fundamental forces; they involve strong, dynamical
spacetime
curvature, supra-nuclear matter densities, relativistic speeds, powerful
neutrino emission and magnetic field production.  They are strong sources
of gravitational waves.  In some cases, they might also produce bright
electromagnetic signals via kilonovae and/or gamma ray bursts and be a
significant site for the production of heavy elements.  Until recently,
they were known to us only by theory, their rate of occurence a matter
of loosely-constrained speculation, with no observational data.  Now
that has changed.

On Jan 5, 2020, during the LIGO O3 observing run, LIGO Livingston registered
a gravitational wave signal,
designated GW200105, consistent with the late inspriral of a BHNS binary.
(LIGO Hanford
was not operational at the time.)  Ten days later, LIGO Hanford, LIGO
Livingston, and Virgo coincidently detected another gravitational wave
signal, GW200115, also consistent with a BHNS
inspiral~\cite{LIGOScientific:2021qlt}.  The two signals were reported
together, although GW200115 was the far more confident detection; in fact,
GW200105 was later downgraded
to a ``marginal candidate''~\cite{KAGRA:2021vkt}.  The other possible BHNS
detections GW190917 and GW191219 were also reported in O3 catalog
papers~\cite{LIGOScientific:2021usb,KAGRA:2021vkt,Zhu:2021jbw}.

The identification
of these binaries as BHNS systems was based entirely on the inferred masses
of the binary components.  From GW200105 was inferred a primary mass of
$M_1=8.9{}^{+1.2}_{-1.5}\,M_{\odot}$ and a secondary mass of
$M_2=1.9{}^{+0.3}_{-0.2}\,M_{\odot}$.  From GW200115 was inferred a primary mass
of $M_1=5.7{}^{+1.8}_{-2.1}\,M_{\odot}$ and a secondary mass of
$M_2=1.5{}^{+0.7}_{-0.3}\,M_{\odot}$.  For GW191219, the masses were
$M_1=31.1^{+2.2}_{-2.7}\,M_{\odot}$, $M_2=1.17^{+0.07}_{-0.06}\,M_{\odot}$;
for GW190917, $M_1=9.7^{+3.4}_{-3.9}\,M_{\odot}$, $M_2=2.1^{+1.1}_{-0.4}\,M_{\odot}$.
The maximum mass of a non-spinning neutron
star $M_{\rm NS,max}$ is somewhere in the range 2--3.2\,$M_{\odot}$, with
the minimum value coming from the maximum inferred masses of observed
pulsars~\cite{Linares:2018ppq,NANOGrav:2019jur} and the maximum value
derived from causality restrictions on the nuclear equation of
state~\cite{Rhoades:1974abc}.  Thus, any compact object more massive than
this is presumably a black hole.  While black holes could exist with masses
below $M_{\rm NS,max}$, there is no known astrophysical way to form them,
so compact objects with mass less than $M_{\rm NS,max}$ are presumed to
be neutron stars.  The waveform also allowed LIGO-Virgo researchers to
estimate a parameter related to the black hole spin, namely the effective
inspiral spin parameter $\chi_{\rm eff}$.  This is a mass-weighted average
of the component parallel to the orbital angular momentum of the Kerr
dimensionless spin parameter $\boldsymbol\chi\equiv {\bf J}/M^2$.  Thus,
$\chi_{\rm eff} = \left(M_1{\boldsymbol\chi}_1 + M_2{\boldsymbol\chi}_2\right)\cdot\hat{L}/(M_1+M_2)$.  This will most likely be dominated by the
black hole's spin. For GW200105,
$\chi_{\rm eff}=-0.01{}^{+0.11}_{-0.15}\,M_{\odot}$,
consistent with zero, while for GW200115,
$\chi_{\rm eff}=-0.19{}^{+0.23}_{-0.35}\,M_{\odot}$, giving a probability of
88\% that the black hole has a spin {\it negatively} aligned with respect
to the orbital angular momentum.

The detections were a matter of excitement for the astronomy community,
because before there had been no confident detections of BHNS systems
through any signal type.  Radio surveys which have identified binary
neutron star systems in our galaxy
have thus far failed to find any black hole-pulsar binaries.  LIGO's O1
and O2 observing runs found no BHNS signals.  The first part of the O3
run turned up only two ambiguous signals:  GW190426, which may
have been a detector artifact, and GW190814, which had a secondary mass
$M_2=2.59{}^{+0.08}_{-0.09}\,M_{\odot}$ and might be a binary black
hole system~\cite{Fattoyev:2020abc,Essick:2020ghc}.  In
the absence of observations, the event rate of BHNS mergers was highly
uncertain.  Population synthesis studies estimate it within
0.1--800\,Gpc${}^{-3}$\,yr${}^{-1}$, and lack of detections in O1 and O2
suggested an upper bound of $<$ 610\,Gpc${}^{-3}$\,yr${}^{-1}$.  Using
two candidate detections in GWTC-3, admittedly very small number statistics,
allowed the LIGO-Virgo-KAGRA collaboration to constrain the rate of BHNS
mergers to between 7.8\,Gpc${}^{-3}$\,yr${}^{-1}$ and
140\,Gpc${}^{-3}$\,yr${}^{-1}$~\cite{KAGRA:2021duu}.  

%The two
%new LIGO detections, admittedly very small number statistics, give
%a merger rate of $45{}^{+75}_{-33}$\,Gpc${}^{-3}$\,yr${}^{-1}$.  Including
%more questionable signals yields a rate of
%$130{}^{+112}_{-69}$\,Gpc${}^{-3}$\,yr${}^{-1}$.  These remain our best
%estimates of the rate of these events.

We should now be optimistic that LIGO-Virgo will detect more BHNS signals
as it approaches its design sensitivity.  Furthermore, a subset of
BHNS mergers produce potentially detectable electromagnetic counterparts,
carrying information about the fascinating postmerger dynamics.
The two most-studied electromagnetic signal possibilities for a
BHNS merger are a {\it kilonova} and a
{\it gamma ray burst} (GRB).

Kilonovae are caused by matter ejected from the
merger site.  As the matter decompresses, it forms unstable nuclei
whose radioactive decays power a potentially detectable thermal IR/visible/UV
signal lasting days.  (See Section~\ref{sec:outflows} below and the
review~\cite{Metzger:2019zeh}.)

GRBs are non-thermal
high-energy emission originating from a highly-relativistic outflow.
In order to achieve the needed Lorentz factors, a large energy must be
released without an accompanying large load of mass.  Jets from the inner
region near the black hole shooting out along the polar directions are
a preferred scenario.  (See Section~\ref{sec:MHD} on the generation of
these jets.)  Short duration GRBs (lasting less than two seconds) are
thought to have their origin in compact binary [BHNS or neutron star-neutron
  star (NSNS)]
mergers~\cite{Nakar2007}, although the detection of kilonovae following
a couple of $\sim$10\,second GRBs suggests that mergers may also sometimes
produce longer-duration GRBs~\cite{Troja:2022yya,JWST:2023jqa}.  Furthermore,
many short GRBs show ``extended emission'' subsequent to the initial
spike~\cite{Norris:2006rw}.  (For reviews of short GRBs,
see~\cite{Nakar2007,Berger:2013jza}.)  GRBs have higher isotropic equivalent
luminosity than kilonovae
($\sim 10^{50}$\,erg s${}^{-1}$ vs $\sim 10^{40}$\,erg s${}^{-1}$), but their
emission is highly collimated, so the observer must be somewhat close to the
axis to see it.  Kilonovae, on the other hand, emit fairly isotropically
and for longer times.  Thus, they have different strengths and weaknesses
as electromagnetic counterparts to a gravitational wave
signal~\cite{Nissanke:2012dj,Bhattacharya:2018lmw}.  The binary neutron
star gravitational wave event GW170817 was detected as both a GRB and
a kilonova~\cite{GBM:2017lvd}.

Other mechanisms for electromagnetic emission
have been considered and will be mentioned in later sections.  Interaction
of the neutron star magnetosphere with the black hole might create
fast radio burst and X-ray transients (see Section~\ref{sec:magnetospheres}).
If the jet must pierce through previously-emitted ejecta, it will produce
a hot surrounding cocoon, producing its own emission and affecting
observations of the jet~\cite{Gottlieb:2017mqv,Lazzati:2019gxx,Gottlieb:2023vuf}.  Outflow
interacting with the surrounding interstellar medium creates radio
emission (see Section~\ref{sec:outflows}).

This chapter is organized as follows.  In
Section~\ref{sec:phases}, we will provide an overview of the inspiral, merger,
and postmerger process.  In
Section~\ref{sec:parameters}, we will explore the space of possible BHNS
binary systems and attempts to use numerical simulations to map properties
of the pre-merger binary to the post-merger outcome.
Section~\ref{sec:ingredients} presents, one piece at a time, the components
of a realistic simulation of a BHNS merger.  Subsequent sections summarize
what has been learned from numerical simulations about the various aspects
and outputs of the mergers:  gravitational waves (Section~\ref{sec:GW}),
signals from the neutron star magnetosphere (Sec.~\ref{sec:magnetospheres}),
the post-merger disk (Sec.~\ref{sec:disks}), production of magnetic fields and
jets (Sec.~\ref{sec:MHD}), and outflows (Sec.~\ref{sec:outflows}).

For other excellent reviews of BHNS systems see
Kyutoku~{\it et al}~\cite{Kyutoku:2021icp} and Foucart~\cite{Foucart2022}. 
Unless otherwise specified, we will use units for which $G=c=1$ throughout
this chapter.
%Symbols for fluid variables will be as follows:
%rest/baryonic density $\rho_0$, temperature $T$, proton fraction $Y_e$,
%specific internal energy $\epsilon$, gas pressure $P$, specific enthalpy
%$h=1+\epsilon+P/\rho_0$.

\section{Phases of a BHNS merger}
\label{sec:phases}

In this section, we outline the evolution stages of a BHNS system.
We introduce the main concepts that will recur throughout the
chapter and provide physical arguments for what we expect simulations to find.

The most likely formation scenario for a BHNS binary begins with a binary
of high-mass main sequence stars in a region of stellar density low enough
that the binary evolves in isolation (a ``field binary'').  If one
star evolves into a neutron star and the other into a black hole--and no
supernova kick disrupts the binary, there will be a BHNS binary.  However,
only BHNS systems that merge within the age of the Universe are interesting
to us.  Probably this requires the binary to undergo a common envelope phase,
leaving the binary sufficiently compact that gravitational radiation
can bring the neutron star and black hole to merger within a Hubble time.

Consider a binary composed of a black hole of mass $M_{\rm BH}$ and a
neutron star of mass $M_{\rm NS}$ in an orbit with semimajor axis $D$.
We expect $M_{\rm BH}$ to be significantly larger than $M_{\rm NS}$,
perhaps $M_{\rm BH}\approx 5M_{\rm NS}$ will be typical.
(See Section~\ref{sec:parameters}.)  The
binary will have a Keplerian orbital angular frequency
$f\approx M_{\rm BH}{}^{1/2} D^{-3/2}$.  It will thus have a
time-varying mass quadrupole moment and so will radiate
gravitational waves.  These waves carry away energy and angular momentum.
The orbital energy of the binary in Newtonian gravity is
$E=-M_{\rm BH}M_{\rm NS}/2D$, and this
orbital energy will decrease at a rate $\dot{E}=-L_{\rm GW}$, where $L_{\rm GW}$
is the gravitational radiation luminosity.  One can then infer
$\dot{D} = dD/dE\dot{E}$.  Gravitational waves cause the eccentricity to
decrease, so by the time the neutron star and black hole are close, the
orbit will be very close to circular.  (BHNS binaries formed by multi-body
gravitational interactions in dense stellar environments such as globular
clusters, on the other hand, might be very eccentric at the time of merger,
but such events are expected to be much rarer than field binary BHNS
mergers~\cite{Tsang:2013mca}.)  The inspiral rate will at first be very
small, but
as $D$ decreases, the binary will orbit faster, increasing $L_{\rm GW}$
and causing the inspiral rate to increase.  This is the inspiral phase.

Either of two effects might terminate the inspiral phase.  First, the
binary separation might reach the innermost stable circular orbit $R_{\rm ISCO}$,
inside of which orbital motion is unstable, and so the neutron star will
plunge into the black hole on an orbital timescale.  If
$M_{\rm BH}\gg M_{\rm NS}$, this separation is that of test particles orbiting
an isolated black hole, which depends only on $M_{\rm BH}$ and on the black
hole's angular momentum ${\bf J}_{\rm BH}$.  In general, the orbital and
black hole spin angular momentum will not be aligned, but we shall consider
this first as the simplest case.  Then $R_{\rm ISCO}=f(\chi)M_{\rm BH}$ where
$\chi=\left|J_{\rm BH}\right|/M_{\rm BH}^2 < 1$ is the dimensionless spin,
and in Boyer-Lindquist coordinates the limits are $f(0)=6$, $f(1)=1$.  In
any case, the plunge radius $R_{\rm plunge}$ can be taken to be of order
$\sim M_{\rm BH}$.

Second, the neutron star might be torn apart by the tidal force of the
black hole.  This will happen when
the tidal force from the black hole exceeds the neutron star's self-gravity.
Both of these forces depend on the neutron star radius $R_{\rm NS}$. 
From Newtonian physics, the gravitational force of a test body with mass $m$
near the surface due to the neutron star's gravity is
$\sim mM_{\rm NS}/R_{\rm NS}^2$
and the tidal force from the black hole (the difference in the black hole's
gravitational forces for test bodies on the sides near and far from the hole)
is $\sim mM_{\rm BH}R_{\rm NS}/D^3$.  Setting the two forces equal we find the
tidal mass transfer radius $D=R_{\rm tidal}$:
\begin{equation}
  R_{\rm tidal} = \left(\frac{M_{\rm BH}}{M_{\rm NS}}\right)^{1/3}R_{\rm NS}
  = M_{\rm BH} Q^{-2/3}C^{-1}
\end{equation}
where we have introduced two crucial dimensionless binary parameters:  the
{\it mass ratio} $Q\equiv M_{\rm BH}/M_{\rm NS}$ and the neutron star
{\it compaction} $C\equiv M_{\rm NS}/R_{\rm NS}$.

Even after the neutron star begins losing mass, it persists inspiraling as a
gravitationally bound compact object with most of its original mass for a
short time
(an orbit or so) longer before being totally disrupted.  It is sometimes
useful to distinguish the beginning of mass transfer off of the neutron star
at $D=R_{\rm tidal}$ from tidal disruption, when the neutron star has been
torn into a spiral swath.  The latter does quickly follow the former, though,
so we will ignore the distinction for the rest of this section.

Tidal disruption would
be expected to have a profound effect on the gravitational wave signal.
By spreading the neutron star matter, the variation of the quadrupole moment
will drop, and the gravitational wave will quickly damp.  We can estimate
this cutoff frequency, using the fact that the frequency of the dominant
(quadrupolar) gravitational wave mode, $f_{\rm cut}$ is twice the orbital
frequency.  Schematically,
\begin{equation}
  f_{\rm cut} \sim M_{\rm BH}{}^{1/2} R_{\rm tidal}{}^{-3/2}
  \sim M_{\rm NS}{}^{1/2} R_{\rm NS}{}^{-3/2}
\end{equation}
i.e. the cutoff frequency is mostly given by the dynamical timescale of the
neutron star.  This is one way that information about the neutron star is
encoded in the waveform~\cite{Vallisneri:1999nq}.  Unfortunately,
$f_{\rm cut}$ comes to $\approx$kHz,
which is well above LIGO's peak sensitivity.

Thus, we have three possible endpoints of the BHNS inspiral.  If
$R_{\rm plunge} > R_{\rm tidal}$, we have a {\it non-disrupting binary}
whose inspiral ends with the neutron star plunging into the black hole
horizon still intact.  If $R_{\rm tidal} > R_{\rm plunge}$, we have a
{\it disrupting binary} whose inspiral ends when the neutron star, still
outside the black hole, is torn apart by the black hole's gravity.  Finally,
there could be marginal systems where $R_{\rm tidal} \approx R_{\rm plunge}$.
This will be the case if
\begin{equation}
  \zeta \equiv f(\chi)Q^{2/3}C \approx 1
\end{equation}
Disruption is expected, then, for $\zeta\lesssim 1$, which will be more likely
for high black hole spin, low mass ratio (i.e. low $M_{\rm BH}$),
and low compaction (in particular, for large $R_{\rm NS}$).

For non-disrupting BHNS binaries, one would not expect residual matter outside
the black hole after the plunge, which eliminates the possibility of most sorts
of electromagnetic signals.  The infalling neutron star will disturb the
spacetime near the horizon, leading to a distinctive ``ringdown'' segment of
the waveform, and one would expect that to be the end of the story.

For disrupting binaries, the neutron star will elongate into a spiral, with
some material expanding outward away from the black hole and some falling
inward toward the black hole.  We will find that, for non-eccentric
BHNS binaries, mass transfer once begun is unstable, and the neutron star
is always completely destroyed in a single tidal disruption event.  Some
of the outgoing material in the spiral has positive energy and is unbound.
This {\it dynamical ejecta}, permanently expelled from the system, can be
quite massive ($10^{-2}$--$10^{-1}\, M_{\odot}$) with asymptotic speeds of
order 0.2--0.3\,$c$.  As this material expands and decompresses, it undergoes
r-process nucleosynthesis, forming heavy nuclei.  Radioactive decays
heat the ejecta, causing it to radiate as a (red) kilonova.

\begin{figure}
  \includegraphics[width=10cm]{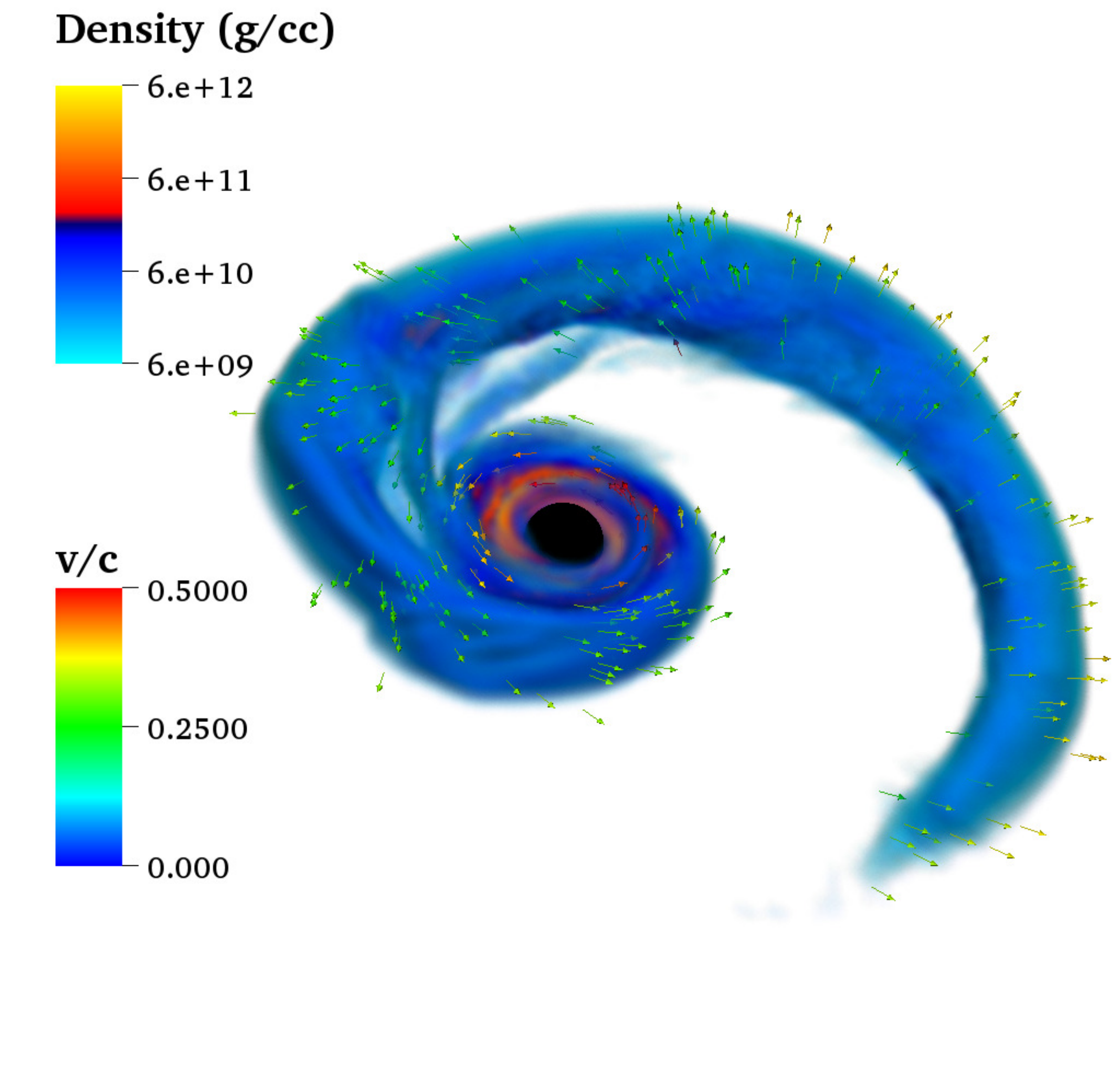} \\
  \includegraphics[width=10cm]{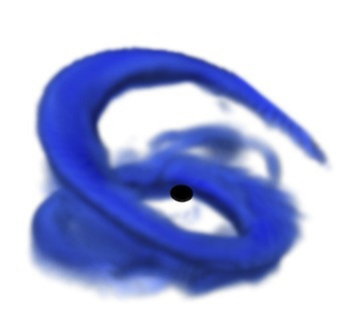}
  \caption
      {
        Tidal disruption converts the neutron star into a spiral swath.
        {\bf Top:}  Incoming matter wraps around the black hole, crashes
        into itself, and forms a hot accretion disk.  Color indicates
        density. 
        (From Foucart~{\it et al}~\cite{Foucart:2014nda}.)
        {\bf Bottom:} In this system with a massive black hole with black
        hole spin $\chi_{\rm BH}=0.9$ misaligned $40^o$ from the orbital
        axis, there is orbital precession, and the stream avoids hitting
        itself. Shown is a density contour. (From
        Foucart~{\it et al}~\cite{Foucart:2012vn}.)
      }
  \label{fig:merger}
\end{figure}

The negative energy matter cannot escape.  Most of the ingoing material falls
immediately into the black hole.  Some of the ingoing spiral has sufficient
angular momentum to circle around the horizon, so that the inflow crashes
into and shears against itself, leading to the formation of an initial
{\it accretion disk}.  (See Fig.~\ref{fig:merger}, top.)  The tidal disruption
and appearance of an orbiting
disk take place in of order a millisecond.  The disk is initially strongly
perturbed, with strong spiral waves.  At the same time, the bound material
that was initially outgoing begins to fall toward the black hole, and this
{\it fallback material} incorporates itself into the disk.  This
{\it post-merger settling phase} lasts for tens of milliseconds, ending with
a roughly axisymmetric black hole accretion system.

Shocks from the merger and settling quickly heat the
gas in the disk to temperatures of a few to 10 MeV.  The maximum
density of the disk shortly after merger can be as high as
$\sim 10^{12}{\rm g}\,{\rm cm}^{-3}$.  At these densities and
temperatures, the gas is efficiently cooled by neutrino emission.  Neutrinos
will carry energy away from the disk and also change the composition of
the gas (the ratio of protons to neutrons), both on timescales of order
10\,ms.  The magnetic field inside the disk is amplified by several
processes, including winding, Kelvin-Helmholtz vortices at shear layers,
and the magnetorotational instability (MRI).  The MRI drives turbulence,
which provides a means of transporting angular momentum outward, causing
the material in the inner disk to accrete into the black hole, the outermost
material to move farther from the hole, and the disk as a whole to spread
out.  The cascade of energy to small scales also provides a heating source
to offset the neutrino cooling.

The subsequent late post-merger evolution is driven by the interplay
of neutrino and magnetic effects.  The transport and heating effects of
magnetoturbulence are often conceptualized and modeled as an effective
viscosity, so that it is common to refer to its effects
as ``viscous'' (e.g. ``viscous winds'', ``viscous timescale''), a custom
we will follow, not without some reservation.  To estimate the timescale
of disk evolution, let us model the transport via a Shakura-Sunyaev alpha
viscosity~\cite{shakura:1973}, in which case the effective kinematic
viscosity is
$\nu=\alpha_{\rm SS} c_s H$, where $c_s$ is the sound speed, $H$ is the disk
thickness, and $\alpha_{\rm SS}$ is a dimensionless constant of order
$10^{-2}$--$10^{-1}$~\cite{Fernandez:2018kax,Hayashi:2021oxy}.  Then the
characteristic timescale for accretion and expansion is
\begin{equation}
  \tau_{\rm visc}\sim \alpha_{\rm SS}^{-1}\Omega^{-1}(H/r)^{-2} \sim 10^2\,{\rm ms}
\end{equation}
where $r$ is the distance from the black hole, and $\Omega$ is the Keplerian
orbital velocity, and we have substituted values typical to a BHNS remnant
disk.

For the first $\sim 100$\,ms, neutrino luminosities are high
(starting at $L_{\nu}\sim 10^{53}$\,erg s${}^{-1}$) and comparable to viscous
heating, and the composition of the fluid is driven to an equilibrium of
charged-current weak nuclear interactions.  Perhaps a percent of the
neutrino luminosity is deposited in the polar regions outside the disk
through the annihilation of neutrinos and antineutrinos, heating matter
there and possibly contributing to the initial formation of a jet.  This
is the {\it neutrino-cooled disk phase}.  As the disk expands, its density
and temperature drop, so that neutrino luminosity decreases, eventually
becoming negligible, so that it no longer balances viscous heating.  The
disk is then said to be in an advective state.  Viscous heating makes the
disk thick and drives disk winds; these winds constitute a second source
of unbound outflow, the {\it disk wind ejecta}, which contribute distinctly
to the kilonova.  This {\it advective disk phase} lasts longer than the
previous phases as the disk depletes. One key goal of BHNS post-merger
simulations is to characterize the disk wind outflows (their mass,
composition, and speed) to determine the expected kilonova signals and
nucleosynthesis output.

The most promising way to produce a relativistic jet, leading to a
GRB, is through a poloidal magnetic field at the polar region
near the horizon by the Blandford-Znajek mechanism~\cite{BZ:1977}.  Thus, a
second key goal of magnetohydrodynamic (MHD) simulations of BHNS post-mergers has been to model
the accumulation of magnetic flux on the horizon, the launching of the jet,
and its eventual loss of power as the disk depletes, in order to enable
comparisons with observed features of GRBs.

As the disk depletes, the accretion rate falls off with time according to
a power law, which would suggest that accretion at a low level at least
might continue for a long time.  Eventually, after of order 100 seconds,
energy from radioactive decays is sufficient to overcome the remnant
disk's binding energy, which will quickly evaporate the disk and put
an end to extended emission~\cite{Lu:2023abc}.

%  Mbh R/D^3 = Mns/R^2
%  C = Mns/R
%  R = Mns/C
%  q = Mbh/Mns
%  Mns = Mbh/q
%  D^3 = Mbh/Mns R^3 = Mbh/Mns Mns^3/C^3
%  D^3 = Mbh Mns^2 C^-3 = Mbh^3 q^-2 C^-3

\section{Binary parameter space and outcome fitting formulae}
\label{sec:parameters}

We expect the outcome of a BHNS merger to depend on $Q$, $\chi$, and $C$.
Let us consider how we expect these parameters to be distributed among
actual BHNS binaries.

Unfortunately, these distributions are very uncertain, especially
regarding the black hole properties.  The black hole can be characterized
by its mass $M_{\rm BH}$ and dimensionless spin ${\bf\chi}$.  Observations
of black holes in low-mass X-ray binaries show a narrow mass distribution
at $7.8{}^{+1.2}_{-1.2}\,M_{\odot}$~\cite{Ozel:2010abc}.  In particular, there
appears to be a mass gap between black hole and neutron star masses, with few
black holes with mass less than $\approx 4.5\,M_{\odot}$~\cite{Farr2011:abc}.
Black holes in binary black holes detected by LIGO-Virgo have broader
distribution extending to higher masses~\cite{LIGOScientific:2020kqk}.
The laws of classical physics permit black holes of any mass, including
within the range of neutron star masses, but no known astrophysical
process would produce black holes with masses below the neutron star
maximum mass.  (Primordial black holes, if they exist, might have low masses,
though.)  From gravitational wave
signals, binary components are often identified as neutron stars or black
holes by their mass, the subtle effects of finite size usually being
undetectable.  Some numerical simulations of BHNS mergers have considered
very low $Q=$1--1.2 BHNS systems, in particular to see if they
can be observationally distinguished from a binary neutron star system
of equal component masses~\cite{Hinderer:2018pei}.

The distribution of inferred $\chi_{\rm eff}$ in binary black holes peaks at
+0.06, with a standard deviation of 0.12~\cite{LIGOScientific:2020kqk}.
It is, of course, possible that the black hole mass and spin distributions
are different for BHNS binaries than they are in X-ray binaries and
black hole-black hole binaries. 
Note that the black hole spin is a vector, and so it enlarges the dimension
of our parameter space by three.  Much numerical work has concentrated on
the case of black hole spin aligned with (i.e. parallel to and in the same
sense as) the orbital angular momentum because this turns out to be the
most optimistic case for tidal disruption and electromagnetic signals.

What about neutron star masses and spins?  Masses inferred from
neutron stars in binaries range from about 1.2 to
2.0\,$M_{\odot}$~\cite{Ozel:2016oaf}.  The mass
distribution of galactic neutron stars in binary neutron star systems is
rather narrowly peaked at 1.33\,$M_{\odot}$, but GW190425, the second
detected binary neutron star merger signals, had a total mass of
3.4\,$M_{\odot}$, implying that higher mass neutron stars in these binaries
are possible~\cite{LIGOScientific:2020aai}.  Furthermore, neutron stars
in binaries with white dwarfs or main sequence stars have a wider distribution
of masses, and peaked at a higher value, than neutron stars in NSNS
systems~\cite{Ozel:2016oaf}.  We have seen that candidate BHNS gravitational
wave signals infer a range of $M_{\rm NS}$ from 1.2 to 2.1\,$M_{\odot}$, with
GW200115, the most confident case, having an inferred neutron star mass of
around 1.5\,$M_{\odot}$.

Neutron stars do spin, but the fastest
observed spins are in the millisecond range (millisecond pulsars), with
a corresponding spin frequency ($10^2$Hz) significantly smaller than the
frequency of a BHNS system at merger ($10^3$Hz).  Furthermore, neutron stars
with magnetic fields spin down and will have plenty of time to do so
during the long inspiral.  On the other hand, tidal forces and neutron
star viscosity are too weak to spin up and tidally lock the neutron star
before merger~\cite{Kochanek:1992abc,Bildsten:1992abc}. 
Therefore, neutron stars are usually approximated
as being irrotational (i.e. a curl-free velocity field) when constructing
initial data for merger simulations. 
However, the effect of a non-trivial neutron star spin has been investigated
in some simulations~\cite{East:2015yea,Ruiz:2020elr}, where it was found that
the neutron star spin can affect the disk and dynamical ejecta masses.

It should be remembered that the total neutron star mass $M_{\rm NS}$
is distinct from the star's ``rest'' or ``baryonic mass'' $M^b_{\rm NS}$.
$M_{\rm NS} < M^b_{\rm NS}$ because of the negative gravitational potential
energy, so $M^b_{\rm NS} - M_{\rm NS}$ is the star's binding energy.

Given the neutron star mass, and assuming slow spin ($\ll$ kHz), the
radius of the neutron star, and hence the compaction, are determined by
the properties of neutron star matter as codified in the equation of state,
which supplies the pressure $P$ as a function of rest mass density $\rho_0$.
This function $P(\rho_0)$ is presumably unique--cold neutron stars have no
free composition variables, unlike main sequence stars and white dwarfs.
Unfortunately, this function is also unknown at high densities.  One of
the driving scientific interests in BHNS and NSNS binaries is to constrain
it.  This can be done by evolving systems with the same binary parameters
(masses and spins) but different assumptions about the equation of state
and determining how the observable outputs (gravitational wave and
electromagnetic) depend on these assumptions.  Thus, for the purposes of
modeling, we must treat the equation of state as if it varied from one
possible system to another.
%One might worry that this would prohibitively
%expand the parameter space to be explored.  After all, doesn't a
%one-dimensional curve have an infinite number of degress of freedom?
%Fortunately, one can build families of equations of state (i.e. functional
%forms for $P(\rho_0)$) with only a few free parameters, which when varied
%can well cover the range of realistic $P(\rho_0)$ curves.
We will have
much more to say about equations of state later (Section~\ref{sec:eos}),
so we defer this discussion until then.  For the rest of this section,
let us simply consider the compaction $C$ to be a free parameter.

Thus, our 6D pre-merger binary parameter space has coordinates
$(M_{\rm BH}, M_{\rm NS}, {\bf \chi}, C)$.  Can we define a parameter
space of the merger outcome?  We needn't be too concerned with the
number of dimensions of this space; it is set by the number of outputs
with which we are interested and does not affect the number of simulations
we need to perform.  Ideally, we would like to be able to express these
output quantities as functions of the binary parameters.  This can be done
by using physical intuition to guess a form of the function, adding free
parameters that can be fit using data from numerical simulations. 
The following outcome quantities have received the
most attention.

\begin{figure}
  \includegraphics[width=12cm]{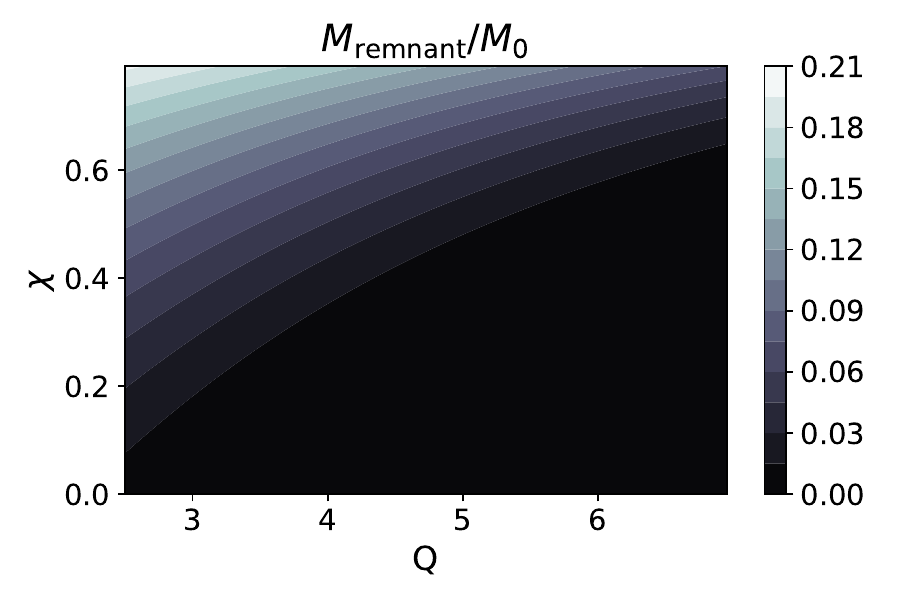} \\
  \includegraphics[width=12cm]{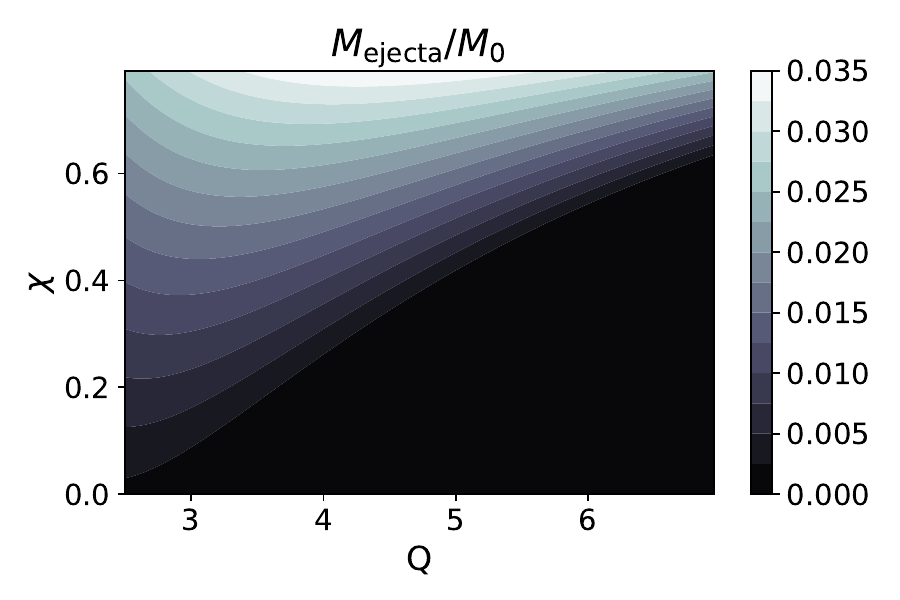}
  \caption
      {Fraction of neutron star baryonic mass surviving in post-merger
        components as a function of mass ratio $Q$ and dimensionless
        aligned black hole spin $\chi$ according to fitting formulae.
        Both plots assume a neutron star with compaction 0.161 and
        specific binding energy 0.094. 
        {\bf Top: } remnant mass after 10\,ms according to the formula of
        Foucart~{\it et al}~\cite{Foucart:2018rjc}.  {\bf Bottom: }
        dynamical ejecta mass according to the formula of
        Kawaguchi~{\it et al}~\cite{Kawaguchi:2016ana}.
      }
  \label{fig:masses}
\end{figure}

A quantity whose dependence on binary parameters has received much attention
is the post-merger baryonic mass, the material that ``survives'' by not
falling immediately into the black hole $M^b_{\rm remnant}$.  Note that this
``remnant'' includes disk, fallback, and dynamical ejecta.  Since the disk mass
is not constant, but
depletes due to accretion and wind, one must specify a time at which the
mass is measured--10\,ms after merger, for example.  From the
discussion in Section~\ref{sec:phases} above, we know that the presence
of tidal disruption (and thus the possibility of a disk) depends on
$Q$ $C$, and $\chi$--the latter through its effect on $R_{\rm ISCO}$.  One
might guess that the remnant mass depends on the same parameters.  In fact,
this works fairly well, as Foucart and collaborators have shown by positing
$M^b_{\rm remnant}/M^b_{\rm NS}$ to be a simple function of the
dimensionless numbers 
$\eta\equiv Q/(1+Q)^2$ (the ``symmetric mass ratio''), $C$, and
$R_{\rm ISCO}(\chi)/M_{\rm BH}$.  They then used the results of available
numerical simulations to fit the free parameters, and are able to achieve
good a reasonably good fit~\cite{Foucart:2012nc,Foucart:2018rjc}.

Fits also exist for the mass and speed of dynamical ejecta.  Kawaguchi
{\it et al}~\cite{Kawaguchi:2016ana} carried out a large number of BHNS
simulations, analyzed dynamical ejecta, and introduced fitting formulae.  Their
ejecta mass function involves $C$, $Q$, $R_{\rm ISCO}/M_{\rm BH}$, and
the neutron star's specific binding energy $1 - M_{\rm NS}/M^b_{\rm NS}$.
(See also~\cite{Kruger:2020gig} for a fit with better behavior for very
compact neutron stars.)  The ejecta mass is highest for $Q\approx 3$;
it is small for very symmetric and very asymmetric
binaries~\cite{Hayashi:2020zmn}.  The ejecta velocity is adequately fit
as a linear function of $Q$ (with higher mass ratios producing faster ejecta).

Looking at the disk and dynamical ejecta masses in various regions of
parameter space (see Fig.~\ref{fig:masses}), we see that
surviving matter tends to be disk-dominated at low mass ratio but
ejecta-dominated at high mass ratios.  Some fraction of the disk,
perhaps between 5\% and 30\% of its mass, will be ejected in winds
(see Section~\ref{sec:outflows} below).  Disk outflow mass will
then presumably also
dominate dynamical ejecta mass only for very low $Q\lesssim 3$.  For some
range of parameters, the two components will both be important, and
for high $Q$ (and high prograde spin, so that the binary is disrupting),
ejecta will mostly be dynamical.  This is important, because we will see
that the two ejecta components have very different properties.

Attempts have also been made to fit outcomes related to the spacetime.
Pannarale fit for
the post-merger black hole's mass and spin~\cite{Pannarale:2012ux,
  Pannarale:2013jua}.  (See also~\cite{Kyutoku:2011vz}.)
The gravitational wave cutoff frequency could
then be fit as a function of these remnant
properties~\cite{Pannarale:2015jka,Pannarale:2015jia}.  We will return
to modeling the gravitational waveform in Section~\ref{sec:GW}.

\section{Ingredients of a numerical simulation}
\label{sec:ingredients}

\subsection{Gravity, plus a bit of history}
\label{sec:gravity}

\subsubsection{Newtonian}
\label{sec:newtonian}

BHNS binaries are strong gravity systems, but for decades numerical
relativists were unable to stably evolve dynamical black hole spacetimes
(although they could make progress on binary neutron
stars~\cite{Shibata:1999wm}),
so the early BHNS simulations were carried out using Newtonian
physics, with the neutron star represented as an ideal fluid and the
black hole as a point mass~\cite{Kluzniak:1997cm,Lee:2000uz,Lee:2001ae,
  Janka:1999qu,Rosswog:2004zx}.  A key question was what happens when
$R_{\rm tidal}$ is reached and
mass transfer begins.  These simulations found cases in which mass transfer
once begun leads to the complete disruption of the neutron star but also
cases where mass loss causes the neutron star to retreat into its
Roche lobe so that the flow of matter off the neutron star stops and a
neutron core survives to
another episode of mass transfer.  They found soft equations of state
to favor unstable mass transfer, stiff equations of state to favor episodic
mass transfer.  In fact this turned out to be an artifact of Newtonian
gravity; relativistic mass transfer is more unstable.

In fact, one can get qualitatively correct results, for which tidal
disruption happens in one pass even for stiff realistic equations of
state, by replacing the Newtonian point mass potential with
``pseudo-Newtonian potentials'' designed to mimic certain features of
black hole orbital dynamics~\cite{Rosswog:2005mf,Ruffert:2010pw}.  The most
common of these are the Paczynski-Wiita potential~\cite{PW:1980} (for
modeling nonspinning black holes) and the Artemova~{\it et al}
potential~\cite{Artemova:1996abc} (for modeling spinning black holes).

It is worth pausing to gain a sense of these techniques because much of the
work on late-time accretion and disk wind ejecta to be discussed later
uses Newtonian physics and pseudo-Newtonian potentials.  The Paczynski-Wiita
potential is just $\Phi(r)=-M_{\rm BH}/(r - 2M_{\rm BH})$.  The simple addition
of the
extra term in the denominator reproduces the innermost stable circular orbit
location of a Schwarzschild black hole (for Schwarzschild radial coordinate).

\subsubsection{Relativistic}
\label{sec:relativistic}

It is beyond the scope of this chapter to explain the techniques used
to stably evolve spacetime metrics (for a book-length treatment,
see~\cite{BaumgarteShapiro:2010}), but perhaps we can briefly convey the flavor of it.
To handle gravity relativistically, one must calculate the spacetime metric
$g_{\mu\nu}$.  The metric provides not only what our Newtonian intuition
regards as gravitational effects but also automatically supplies the
properties of the black hole and of the gravitational waves, because these
three are all just different aspects of the spacetime curvature.  The
Einstein field equations are nonlinear second-order partial differential
equations for $g_{\mu\nu}$.  The field equations come in two types:
{\it evolution} equations, which supply the time derivatives that tell how the
metric changes from one time slice to the next, and {\it constraint}
equations, which restrict how the fields must relate at each time.  This
is entirely analogous to the split in Maxwell's equations, e.g. Faraday's
law $\partial_t{\bf B}=-{\bf \nabla\times E}$ is an evolution equation
while Gauss's Law $\nabla\cdot E=4\pi\rho$ is a constraint which must
be satisfied at each time.  (Analytically, if the constraints are satisfied
at one slice of time, the evolution equations will keep them satisfied at
later times, as with Maxwell's equations.  Whether numerical-error induced
constraint violations remain small depends on one's formulation of Einstein's
equations.)  The numerical relativist must
supplement Einstein's equations with a third set of equations, the
{\it gauge} conditions, which determine the choice of coordinates, e.g.
how spacetime is to be sliced into space and time.

A key part of any numerical relativity calculation is the construction of
appropriate initial data.  For a BHNS simulation, this must correspond to
a black hole and a neutron star in circular orbit.  Numerical costs
(above all the number of timesteps involved) motivate us to create binaries
in the late inspiral stage, no more than roughly a dozen orbits before
merger.  To be a time slice of a solution of Einstein's equations, the
constraint equations must be satisfied at
this $t=0$ slice.  This amounts to four elliptic partial differential
equations that must be solved.

In addition, one wants to impose the
conditions that the star is in equilibrium and the binary is in circular
orbit, which will not be true for just any choice of density distribution,
orbital rotation rate, and (constraint satisfying) metric.  It is impossible
to impose an exact, consistent circular orbit symmetry on the spacetime,
%(i.e. one
%where $\mathcal{L}_{\partial_t + \Omega\partial_\phi}g_{\mu\nu}=0$),
impossible because a BHNS orbiting in perfect circles forever 
is not a solution to Einstein's equations--actually, there must be
gravitational radiation and inspiral.  What is actually done, in
what is called the extended conformal thin sandwich (XCTS) formulation,
is to freely choose some metric variables and then use the constraints
and assumed circular orbit to solve for the remaining five.  For example,
one factors the three-metric as $\gamma_{ij}=\Psi^4\tilde{\gamma}_{ij}$,
chooses $\tilde{\gamma}_{ij}$, e.g. to be flat or to resemble a single
black hole
space, and solves for $\Psi$ with a constraint equation.  Making slightly wrong
assumptions makes the problem tractable, but it comes at a cost.  An
incorrect $\tilde{\gamma}_{ij}$ means unphysical (``junk'') gravitational
radiation in the initial data.  Ignoring inspiral produces slightly eccentric
orbits (which can be reduced by adding some initial
inspiral~\cite{Pfeiffer:2007yz}).

Equilibrium must also be imposed on the fluid.  The Euler equation can
be integrated to give an algebraic condition (with integration constant),
the relativistic generalization of the Bernoulli integral.  This turns out
to be an equation for $h\Gamma$, where $h$ is the specific enthalpy, and
$\Gamma$ is the Lorentz factor.  Thus, given the fluid velocity, equilibrium
gives $h$ and thus (via the equation of state) $\rho_0$.  The condition that
the star is non-spinning is translated as the condition that the flow is
irrotational.  Recall that in Newtonian physics, this means
${\bf\nabla\times v} = {\bf 0}$ so that ${\bf v}\equiv{\bf\nabla}\phi$,
and the relativistic version is similar.  An elliptic equation for the
velocity potential $\phi$ is supplied by the continuity equation.

To the extent that the inspiral timescale is much longer than the orbital
timescale, one can think of the binary as simply evolving from one of these
quasi-equilibrium circular orbit states to another.  One can, then, track
the inspiral evolution of a single binary by creating a sequence of these
equilibria at different separations, holding fixed quantities like the
neutron star baryonic mass and black hole irreducible mass (expected to be
nearly conserved since little of the gravitational wave energy is swallowed
by the black hole.)  At smaller separations, the orbital angular velocity
$\Omega$ increases; in fact, $\Omega$ is a better parameter along the
sequence than coordinate separation, because it can be calculated in
a coordinate-invariant way.  As the binary inspirals to higher $\Omega$,
the total energy (ADM mass) of the binary goes down, with the reduction
presumably accounted for by gravitational wave emission. 
Before numerical relativity evolutions were possible,
these sequences provided our best understanding of the late
inspiral~\cite{Miller:2001zy,Grandclement:2006ht,Taniguchi:2006yt}.  The
most interesting question was how the quasi-equilibrium sequences end.
The ISCO of a sequence is where the binary's energy as a function of $\Omega$
reaches a minimum.  The onset of tidal mass transfer is indicated by the
formation of a cusp shape on the stellar surface.  Unfortunately, this final
and most interesting part of the sequence is precisely where inspiral is
becoming rapid and the quasi-equilibrium assumption certainly breaks down. 
Thus, we are compelled to do full evolutions to model the end of the inspiral
credibly.

For further reading on initial data construction, see~\cite{Tichy:2016vmv}.

Numerical relativity experienced a breakthrough in its ability to stably
evolve binaries with black holes in 2005.  (For a review of numerical
relativity centering on this breakthrough and the work it enabled,
see~\cite{Duez:2018jaf}.)  The next year,
Loffler~{\it et al}~\cite{Loffler:2006abc}
simulated a BHNS head-on collision, and Shibata and Uryu carried out the first
fully relativistic simulations of BHNS merger starting from roughly circular
orbit~\cite{Shibata:2006bs}.  This was quickly followed by other
groups~\cite{Etienne:2007jg,Duez:2008rb,Chawla:2010sw}.

\subsection{Equation of state}
\label{sec:eos}

BHNS simulations must model matter in a wide range of densities and
temperatures.  The high density matter inside neutron stars and the
low density matter in disks and outflows each present their own challenges.

In general, the gas will be composed of a combination of photons, leptons,
and baryons, with the baryons divided into free protons ($p$), free
neutrons ($n$), and
many different species of nuclei.  At temperatures above about 0.5\,MeV,
we can invoke the wonderful simplification of {\it nuclear statistical
  equilibrium} (NSE).  That is, strong nuclear reactions proceed so quickly that
the abundances of each nuclide (including free nucleons) quickly come to
their equilibrium values, and this equilibrium is maintained as fluid
elements evolve.  Then the composition of each isotope is fixed by a
network of Saha-like equations, and the state of the fluid can be described
by only three variables:  the number density $n$ (actually, the numerical
relativity community prefers the equivalent number, the baryonic density
$\rho_0$, which is just $n$ multiplied by a standard baryonic mass), the
temperature $T$, and the fraction of baryons that are protons
$Y_e\equiv n_p/(n_p+n_n)$.  The latter, $Y_e$, is sometimes called the
reduced electron fraction and is what is being referred to when numerical
relativists speak of the ``composition'' of our outflows.  Note that
$Y_e$ of a fluid element can only change because of charged-current weak
interactions (which can change protons to neutrons, and vice versa), which
also create and absorb neutrinos.  The timescale for these weak interactions
to come to equilibrium is not always small compared to evolution timescales,
so $Y_e$ must be evolved.  The equation of state thus provides pressure $P$
and specific internal energy $\epsilon$ as functions of $(\rho_0,T,Y_e)$.
Nuclear physics equation of state models provide these to numerical
relativity codes in tabulated form.

Before merger, the neutron star matter is very degenerate, so one can
take $T=0$.  Furthermore, the neutron star will have had time to settle
to equilibrium to weak interactions (``neutrinoless beta-equilibrium'').
The latter
equilibrium condition ($\mu_p+\mu_{e^-}-\mu_n=\mu_{\nu_e}=0$, where
$\mu_X$, denoting the chemical potential of particle X, is a function
of $\rho_0$, $T$, and $Y_e$ given by the equation of state) amounts to
a condition on $Y_e$, so the equation of state becomes one-dimensional:
$\epsilon=\epsilon(\rho_0)$, $P=\rho_0^2d\epsilon/d\rho_0$.
%d(e/r) = -p d(1/r)

This 1D equation of state is all that is needed for initial data and,
because most of the gravitational waveform comes from the inspiral, it
is sufficient for most gravitational wave studies.  Systematic studies
of the effect of equation of state have been carried out assuming the
piecewise polytrope family of equations of state.  Here, one divides the
density into intervals and takes the pressure in each interval to
be a simple power law of the density.  For example, in interval $i$
covering densities between $\rho_{0,i-1}$ and $\rho_{0,i}$, one has
$p(\rho_0) = \kappa_i\rho_0^{\Gamma_i}$, where $\kappa_i$ and $\Gamma_i$
are constants.  For chosen $\rho_{0,i}$, $\Gamma_i$, and $\kappa_0$,
one can infer the other $\kappa_i$ by requiring continuity in $p(\rho_0)$.
The range of realistic behavior can be captured with three free
parameters~\cite{Read:2008iy}.  One downside to piecewise polytropes
is that they are not smooth at the boundaries of intervals, meaning
the sound speed is discontinuous.  There is a simple generalization
to the equation of state family that fixes this~\cite{OBoyle:2020qvf}.

An alternative flexible but smooth equation of state family is available
using ``spectral'' equations of state~\cite{Lindblom:2010bb,Foucart:2019yzo}.
Here one takes the independent
variable to be $x\equiv\ln(\rho_0/\rho_0^{\rm ref})$ for some reference
density $\rho_0^{\rm ref}$.  The adiabatic index $\Gamma(x)\equiv d\ln(P)/dx$
is specified as a sum of basis functions (hence ``spectral''), and then
$P(x)$ and $\epsilon(x)$ can be recovered by appropriate integrals.
In fact, the simple choice of a polynomial $\Gamma(x)=\Sigma_{n=0}^Na_nx^n$
suffices.

After merger, the densities in disks and outflows are low enough that
nuclear forces are unimportant (except in the sense of holding nuclei
together), and the equation of state is just a combination of classical
ideal cases for free nucleons and nuclei, ideal Fermi gases for
electrons/positrons, and photon radiation.  As the disk expands, the
NSE equilibrium changes from free nucleons to alpha particles and
heavy nuclei.  This recombination of nucleons releases binding energy,
providing a source of thermal energy that is important for disk winds.
Note that this is a reversible process; it does not generate entropy.
Conservative MHD codes do not need to add a ``heating'' term, since the
negative binding energy automatically means more thermal energy at a
given total energy.

Simulations that follow disks and outflows for second timescales run into
the problem of temperatures $\ll 0.5$\,MeV, which makes the continued
assumption of NSE dubious.  In particular, heating from r-process
nucleosynthesis might  on second timescales provide a significant source
of thermal pressure, and such an effect
cannot be captured with an NSE code.  Unfortunately, the only adequate
solution of dropping NSE and
evolving isotope abundances via nuclear reaction rates would mean an
explosion of evolution variables describing the composition of the fluid.
Instead, a few studies have tried to add phenomenological heating terms
to estimate the effects of this heating on outflows and
fallback~\cite{Metzger:2010abc,Rosswog:2013kqa,Desai:2018rbc}.  More
recently, the necessary step of dropping NSE has been undertaken with
simulations with a nuclear reaction network evolving nuclide
abundances coupled to 2D ray-by-ray hydrodynamic evolution of the
outflow~\cite{Magistrelli:2024zmk}.

\subsection{Neutrino transport}
\label{sec:neutinos}

After merger, the surviving orbital material heats to $T\sim$MeV, and
neutrinos are copiously produced by weak nuclear interactions, including
particularly the charged-current reactions, which alter $Y_e$ and produce
electron-flavor neutrinos and antineutrinos
\begin{eqnarray}
  \label{eq:epnnu}
  e^- + p   &\rightarrow& n + \nu_e \\
  \label{eq:enpnu}
  e^+ + n   &\rightarrow& p + \overline{\nu}_e
\end{eqnarray}
and the electron-positron pair annihilations, which produce neutrinos
of all flavors $i$
\begin{equation}
  e^- + e^+ \rightarrow \nu_i + \overline{\nu}_i
\end{equation}
Immediately post-merger, neutrino luminosities reach
$L_{\nu}\sim 10^{53}$--$10^{54}$\,erg s${}^{-1}$.  The thermal energy of the
newborn disk is $E_T\sim 10^{52}$erg, suggesting a thermal timescale
of $\tau_{\rm th}\sim 10$--$100$\,ms~\cite{Deaton:2013sla}.

Electron-type neutrinos can also be absorbed by the reverse of the processes of
Eq.~\ref{eq:epnnu} and Eq.~\ref{eq:enpnu}.  In addition, neutrinos of all
flavors scatter off of free nucleons and nuclei.  There will thus be,
for neutrino species $i$, an
opacity for absorption $\kappa_{i,a}$, an opacity for scattering
$\kappa_{i,s}$, and
a total opacity $\kappa_i=\kappa_{i,a}+\kappa_{i,s}$.  (As with photon transport,
$\kappa_i$ is neutrino energy-dependent, and absorption and scattering can have
somewhat different effects--details which are important in a radiation
transport code but which will be ignored henceforth in this overview.)
The associated mean free path is $\ell_i = 1/\kappa_i$ [or
  $1/(\kappa_i\rho_0)$, depending on how one defines $\kappa_i$].  How important
will these scatterings and absorptions be in a post-merger disk?
To answer that, one can compute the optical depth, which for a disk of
height $H$ will be $\tau_i\sim H/\ell_i$.  If $\tau_i\ll 1$, the disk is
optically thin, i.e. transparent, and opacity is unimportant.  Emitted
neutrinos travel outward from their emission sites along almost null
geodesics.  If $\tau_i\gg 1$, the disk is optically thick; neutrinos are
trapped, come to (Fermi-Dirac) equilibrium, escape more slowly by random
walk / radiative diffusion, and can even be advected with fluid into the
black hole.  For BHNS disks, $\tau_i$ is at most of order $\sim 10$, but after
a viscous timescale of disk depletion will almost certainly be optically
thin.  

Because $\tau_i$ ranges from low to high, evolving the neutrino fields is
difficult.  Each neutrino species is described by a distribution function
(the density in phase space) $f({\bf x},{\bf p})$.  The evolution equation
of $f$, the Boltzmann transport equation, is simple, but the fact that it
is 6-dimensional makes it expensive even to store a reasonable numerical
representation of it.

Many of the BHNS simulations that include neutrino
effects use {\it neutrino leakage}, i.e. they do not evolve the neutrino
fields at all, but only add a cooling
term in the fluid evolution, as well as a source term for $Y_e$ to account
for lepton number changes from Eq.~\ref{eq:epnnu} and Eq.~\ref{eq:enpnu}.
For $\tau_i\ll 1$, one can get these from the local emission rates;
for higher $\tau_i$, one must approximate, at least to order of magnitude,
the effect of diffusion.  These methods decently capture neutrino cooling
and were used for the earliest studies of BHNS mergers that incorporated
weak interactions~\cite{Janka:1999qu,Rosswog:2005mf,Deaton:2013sla,
  Foucart:2014nda,Brege:2018kii}, but they
have the serious drawback that
neutrinos emitted in one part of the grid cannot be absorbed in another
part.

A more sophisticated way to model neutrinos is
provided by an {\it M1 moment closure scheme}~\cite{Thorne:1981nvt,Shibata:2011kx},
which rather than evolving all of $f$ only involves
its lowest two moments:  the neutrino energy and momentum densities.  Most
codes integrate the specific neutrino energy and momentum density over
neutrino energies and evolve only the energy-integrated densities.  Such
transport schemes are called ``grey''.  The evolution equation of each
neutrino moment involves the next higher moment, so in order to obtain
a closed, finite set of equations, one must assume the form of the first
moment not evolved, i.e. one must provide a ``closure condition''. 
M1 does capture neutrino absorption, and it has been used in two studies
of BHNS mergers~\cite{Foucart:2015vpa,Kyutoku:2017voj}, which found that
neutrino irradiation of dynamical ejecta does not significantly affect
its composition, but the disk $Y_e$ profile is noticeably different than
predicted by leakage.  Although
a great improvement upon leakage, grey M1 remains imperfect, both because of
imperfection in the closure condition and the sacrifice of spectral information.

Recently, some numerical relativity codes have evolved the full Boltzmann
transport equation for the full neutrino distributions using {\it Monte Carlo}
methods~\cite{Miller:2019gig,Foucart:2021mcb}.  Monte Carlo transport
coupled to MHD was used to evolve a post-merger-like disk with the code
$\nu${\tt bhlight}~\cite{Miller:2019dpt}.  The disk had an initial optical
depth of $\sim 10$, so neutrino absorption is able to significantly affect
the $Y_e$ distribution of the disk and its outflows.

In addition to carrying energy, emitted neutrinos carry their momentum
away with them, meaning neutrinos can provide a source of physical viscosity.
Under some conditions, this viscosity could significantly slow the growth
of the MRI~\cite{Masada:2008nb}.  If the MRI is suppressed, momentum transport
in the disk will be greatly slowed, since the neutrino viscosity itself would
be much weaker than the turbulent effective viscosity that would otherwise
be present. 
Guilet~{\it et al}~\cite{Guilet:2014kca} have made a careful study of the
effect of neutrino momentum transport on the MRI.  They find that neutrino
effects behave like a viscosity if the neutrino mean free path $\ell$ is
much smaller
than the wavelength of the fastest growing MRI mode:
$\ell < \lambda_{\rm MRI}$.  If $\ell > \lambda_{\rm MRI}$, neutrinos act as
a drag force.  For realistic (i.e. not-magnetar) initial fields, early
post-merger $\lambda_{\rm MRI}$ is small enough that neutrinos will act as
a drag.  Simulations indicate that the drag's damping timescale is at least
initially much longer than the orbital timescale on which the MRI
grows~\cite{Foucart:2015vpa,Kiuchi:2015qua}, meaning neutrinos probably
will not prevent a phase of exponential field growth.

Adequate treatment of neutrino physics remains a serious challenge for
BHNS simulations.  Even the most sophisticated of the above schemes ignore
neutrino flavor oscillations, which might affect outflow
compositions~\cite{Malkus:2015mda,Li:2021vqj}.  For a review of
neutrino transport methods in numerical relativity, see~\cite{Foucart:2022bth}.

\section{Gravitational waves}
\label{sec:GW}

The inspiral gravitational waveform of a BHNS binary is, like the waveform
from a binary black hole (BHBH) or a NSNS binary,
well described by the waveform expected for
a binary of two point masses.  However, the fact that the neutron star is
an extended object does leave a small imprint on the waveform.  The
gravity of the black hole tidally deforms the neutron star, giving it a
quadrupole moment which affects the binding energy $E$ and contributes to
the gravitational wave luminosity $L_{\rm GW}$.  To lowest post-Newtonian
order, the effect of tides on the gravitational wave phase depends only
on one numerical property of the neutron star:  its tidal deformability
$\Lambda$, the ratio of the induced quadrupole moment of the star to
the perturbing tidal gravitational field, in this case from the black
hole~\cite{Flanagan:2007ix}.
This number $\Lambda$ is strongly dependent on the compaction of the neutron
star; it is, in fact proportional to $R_{\rm NS}{}^5$.  Thus, if this
phase drift could be detected, it would provide a way of estimating the
neutron star radius.  Unfortunately, the tidal effect is smaller at higher
$Q$, so BHNS systems are less promising for this sort of measurement than
NSNS systems.

\begin{figure}
  \includegraphics[width=10cm]{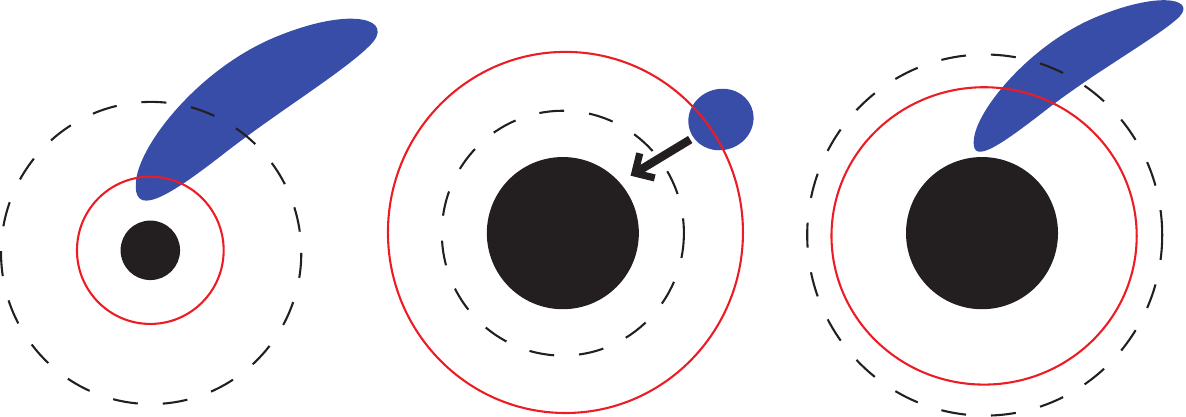} \\
  \includegraphics[width=10cm]{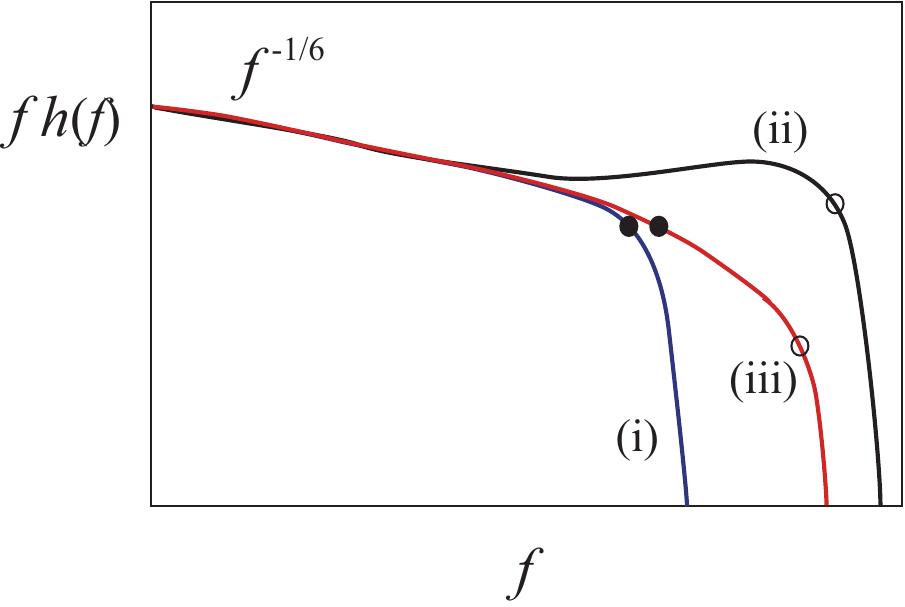}
  \caption
      {Schematic pictures of three different merger types and their
        corresponding gravitational wave spectra.  In the top pictures,
        the solid circle is
        the ISCO location, the dashed circle the location of tidal
        disruption.  The types are (i) disruption with disrupted star
        larger than the black hole, (ii) plunge, and (iii) disruption
        with disrupted star smaller than the black hole.
        Reprinted figure with permission from Kyutoku, Okawa, and Shibata,
        Phys. Rev. D {\bf 84} 064018 (2011)~\cite{Kyutoku:2011vz}.
        Copyright 2011 by the American Physical Society.
      }
  \label{fig:bhnswaves}
\end{figure}

The effect of the end of the inspiral on the gravitational wave is dramatic,
although unfortunately in the poorly accessible kHz range.  The best way to
understand the BHNS waveform is to compare it to a BHBH waveform.  BHBH
waveforms have three parts:  (1) the inspiral waveform, during which the
amplitude and frequency ramp up, (2) the merger waveform, at which the
waveform amplitude peaks, and (3) the ringdown, a damped oscillation waveform
representing the decaying perturbed modes of the settling black hole.
Recall that BHNS inspiral may end, and merger may commence, in one of three
ways:  plunge ($R_{\rm ISCO} > R_{\rm tidal}$), tidal disruption
($R_{\rm ISCO} < R_{\rm tidal}$), or both together
($R_{\rm ISCO} \approx R_{\rm tidal}$).  Using numerical relativity
simulations for a range of $Q$, Shibata~{\it et al.}~\cite{Shibata:2009cn}
found that these three types of inspiral termination correspond to three
types of BHNS merger gravitational waveforms, as illustrated in
Figure~\ref{fig:bhnswaves}.

Systems in which the neutron star plunges into the black hole before
significant mass transfer have waveforms very similar to BHBH waveforms,
with inspiral,
merger, and ringdown.  (Indeed, the waveform is almost identical to that of a
BHBH system with the same component masses~\cite{Foucart:2013psa}.)

If tidal
disruption happens well outside $R_{\rm ISCO}$, the waveform cuts off
during the inspiral segment, and the wave amplitude
decreases quickly as the disrupted neutron star spreads out and the binary's
quadrupole moment drops.  The matter falls through a broad, roughly
axisymmetric region of the black hole and does not excite strong, coherent
quasi-normal modes on the black hole, so in this case there is no ringdown
waveform.  One would guess that the frequency at which the waveform drops
would be given by $f_{\rm cut}$ (cf. Section~\ref{sec:phases}), the frequency
corresponding to where mass transfer begins.  Actually, the true cutoff
frequency is higher, at the frequency at which the neutron star is fully
tidally disrupted.  The distinction has a significant effect on the
dependence of the gravitational wave cutoff on the neutron star compaction~\cite{Kyutoku:2010zd,Kyutoku:2011vz,Kyutoku:2021icp}.  More compact neutron
stars not only begin mass transfer closer to their black hole; they also
survive mass transfer longer.

There remains the intermediate case in which plunge and tidal disruption
happen together.  The type of waveform in this case depends on the black
hole mass.  For $Q\lesssim 3$, one sees an intermediate waveform type
with inspiral and merger but very weak ringdown component.  For higher
$Q$ with tidal disruption near the ISCO, which requires high prograde
black hole spin, the ringdown component remains present.  The reason
is that although the neutron star is spread out by tidal disruption,
if the black hole horizon is bigger than the original neutron star, the
disrupted star still flows through a localized region of the horizon.  

Since $R_{\rm tidal}$ depends on $C$, the waveform cutoff in disrupting
cases has information about the neutron star and its equation of state.
The information contained in the cutoff frequency in fact turns out to
be nearly equivalent to $\Lambda$, presumably because both $\Lambda$ and
$C$ depend (for known $M_{\rm NS}$) mostly on $R_{\rm NS}$.
An ambitious numerical exploration of the BHNS parameter space was
carried out with the {\tt SACRA} code by
Lackey~{\it et al}~\cite{Lackey:2013axa}:  134 simulations varying
mass ratio, (aligned) black hole spin, and neutron star equation
of state.  To explore possibilities of the equation of state, a
piecewise polytrope family with two free parameters was used.  Although
their equation of state space was 2D, a Fisher matrix analysis confirms
that waveforms allow one best to measure one equation of state parameter,
the tidal deformability.

It would be impossible to produce numerical relativity waveforms for every
set of binary parameters that gravitational wave observatories might need
as templates for detection and parameter estimation.  Instead, the hope
is to use
waveforms from a smaller number of BHNS simulations to calibrate some
simpler waveform model for which cases can be generated quickly.  Thus,
finite-size corrections have been added to the
phenomenological (``Phenom'') and effective one-body (``EOB'') families
of binary inspiral-merger-waveforms.  The Phenom BHNS
waveforms~\cite{Lackey:2013axa,Pannarale:2015jka,Thompson:2020nei}
were calibrated to the 134 {\tt SACRA} waveforms.  The EOB BHNS waveform
models~\cite{Matas:2020wab} were calibrated to {\tt SACRA} and {\tt SpEC}
numerical waveforms.

One way to test the adequacy of these models for LIGO-Virgo purposes is
to insert numerical relativity waveforms into LIGO noise and see if
the signals can be detected and the binary and equation of state parameters
correctly
extracted using the model waveforms as templates.  Waveform models are
considered good enough if the systematic errors from model inaccuracies
are lower than the expected statistical errors (given whatever signal to
noise ratio one is hoping for).  See~\cite{Chakravarti:2018uyi,Huang:2020pba}
for checks of this sort.

\section{Magnetospheres}
\label{sec:magnetospheres}

Neutron stars observed as pulsars have magnetic fields that extend outside
the neutron star surface into a magnetosphere region.  In a BHNS binary,
this neutron star magnetosphere could extend to and begin interacting with
the black hole prior to merger.  Electromagnetic signals from this
magnetosphere interaction are interesting for two reasons.  First of all,
the signal could actually precede the merger and so perhaps appear shortly
before a GRB.  Second, this magnetosphere interaction and
accompanying signal could still occur even for nondisrupting BHNS binaries
(recall: $R_{\rm ISCO}>R_{\rm tidal}$), which otherwise seem to be such
duds, electromagnetically speaking.

In the neutron star magnetosphere, the energy density and stresses of
the electromagnetic field dominate over that of the fluid, i.e.
$T^{\mu\nu}_{\rm EM} \gg T^{\mu\nu}_{\rm fluid}$; in this
limit, the fluid's pressure and even its inertia
can be ignored~\cite{Glodreich:1969abc}.  Energy-momentum conservation
then gives
\begin{eqnarray}
  0 &=& \nabla_{\nu}(T^{\mu\nu}_{\rm EM}+T^{\mu\nu}_{\rm fluid})
  = \nabla_{\nu}(T^{\mu\nu}_{\rm EM}) = F^{\mu\nu}\mathcal{J}_{\nu} \\
  \label{eq:ff}
  {\bf 0} &=& \rho_e{\bf E} + {\bf J\times B}
\end{eqnarray}
where $F^{\mu\nu}$ is the Faraday tensor, $\mathcal{J}_{\nu}$ the
4-current, $\rho_e$ the charge density, and ${\bf J}$ the 3-current.
This is called the ``force free limit'' because the above stress-energy
condition is really a statement that momentum doesn't get transferred
from the electromagnetic field to the fluid (i.e. no force on it).  It should be
emphasized that, even though $T^{\mu\nu}_{\rm fluid}$ is negligible, the
force-free limit is very different from the vacuum Maxwell limit.  The
fluid contributes not inertia but free charges which provide conductivity.  Notice
from Eq.~\ref{eq:ff} that ${\bf E\cdot B}=0$.  That is, there can be
no electric potential difference along a magnetic field line; magnetic
field lines are like {\it wires}.

Force-free evolutions often lead to the formation of regions in which the
force-free approximation (and ideal MHD itself) breaks down.  This will happen
at current sheets, where {\bf J} becomes large and the force-free condition
$B^2>E^2$ breaks down.  Inside current sheets, the effects of resistivity
(ignored by ideal MHD) become important, and magnetic field lines reconnect,
releasing energy and producing isolated magnetic field loop regions
(plasmoids).

\begin{figure}
  \includegraphics[width=10cm]{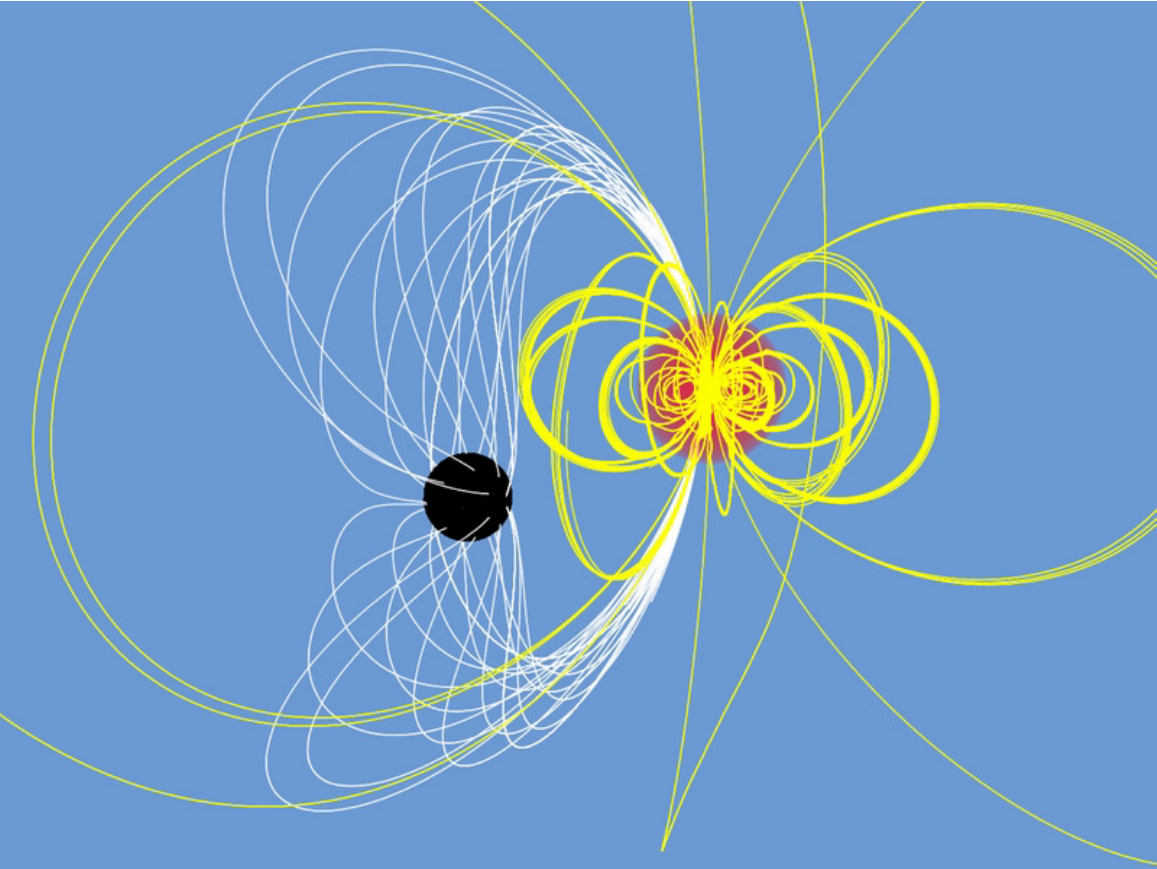}
  \caption
      {Magnetic field lines from a neutron star magnetosphere interacting
        with a black hole before merger.  Reprinted figure with permission
        from Paschalidis, Etienne, and Shapiro, Phys. Rev. D {\bf 88},
        021504 (2013)~\cite{Paschalidis:2013jsa}.  Copyright 2013 by the
        American Physical Society.
      }
  \label{fig:magnetosphere}
\end{figure}

The same unipolar inductor model that has been used to explain the
Blandford-Znajek effect was used to predict electromagnetic energy release
from BHNS magnetosphere interaction~\cite{McWilliams:2011zi}.  In the
original Faraday disk, a spinning conducting disk threaded by a magnetic
field has an electric field (by the MHD condition ${\bf E}={\bf B\times v}$)
and so a potential difference; by touching the disk at different points with
wires, the disk can be used as a battery.  In the Blandford-Znajek mechanism, the black hole
horizon and magnetic field lines play analogous roles.  A similar unipolar
inductor model of a BHNS system would predict an energy-releasing black hole
battery circuit driven by the orbital motion of field lines at the horizon.

This has been modeled in numerical relativity by
Paschalidis~{\it et al}~\cite{Paschalidis:2013jsa} using an ideal MHD
code able to extend to the force-free limit~\cite{Paschalidis:2013gma} and by
East~{\it et al}~\cite{East:2021spd} and Most~{\it et al}~\cite{Most:2023unc}
using resistive MHD codes whose chosen
form of Ohm's law recovers the force-free limit in the appropriate
regime~\cite{Palenzuela:2012my}.  A snapshot from one such simulations
showing magnetic field lines from the neutron star magnetosphere under
the influence of the black hole is shown in Figure~\ref{fig:magnetosphere}.
All of these studies report the electromagnetic energy outflow as a
component of energy output.  This can be calculated from the Poynting flux
through an arbitrary distant sphere.  One can also compute energy drained
from the magnetic field via reconnection in current sheets~\cite{Most:2022ojl}.

The unipolar inductor model is found to
estimate well the Poynting power output near the time of merger.  In the
magnetosphere of an isolated pulsar with spin angular frequency $\Omega$,
a crucial role is played by the light cylinder, the cylindrical radius
$\varpi\approx c/\Omega$ at which the magnetosphere would have to orbit
at the speed of light to spin at the same rate as the star, where one
sees a switch from corotating closed field lines inside to open field
lines and a current sheet.  Similarly, in the BHNS, there is an orbital
light cylinder at $\varpi\approx c/\Omega$ with $\Omega$ the orbital
angular frequency, and the black hole twisting field lines close to this
cylinder plays an important role in the energy release.  Energy is released
both by reconnection heating and by outgoing Poynting flux.  The resulting
radiation might be observed as a fast radio burst or X-rays.

\section{Disk formation and structure}
\label{sec:disks}

Relativistic simulations confirm that tidal disruption occurs in a single
pass, even for stiff realistic equations of state~\cite{Duez:2009yy}.
Episodic mass transfer, as well as other interesting dynamics, {\it is}
possible in general relativity for the merger of {\it eccentric} BHNS
binaries~\cite{East:2011xa}.  Relativistic simulations are also the most
reliable for studying the effect of black hole spin.  Aligned prograde
spin can make what would have been nondisrupting binaries disrupting and
leads to more massive remnant disks~\cite{Etienne:2008re,Kyutoku:2011vz,
  Lovelace:2013vma}, retrograde spin has the opposite effect, and
misaligned spins have intermediate effect.  This influence on
disk mass is reproduced by Foucart's fitting
formulae~\cite{Foucart:2012nc,Foucart:2018rjc} through the
effect of black hole spin on $R_{\rm ISCO}$.  For high mass ratios
$Q\gtrsim 5$, corresponding to what are likely the most common black hole
masses, disruption and massive disk formation are only possible for high
($\chi_{\rm BH}\gtrsim 0.7$) prograde spin~\cite{Foucart:2011mz}, suggesting
the depressing (to multimessenger astronomers and MHD numericists) likelihood
that most BHNS mergers in nature are non-disrupting.  
High $Q$ BHNS binaries that do disrupt have a larger fraction of the
surviving matter outside the black hole in dynamical ejecta as opposed to
disk~\cite{Foucart:2012vn,Kyutoku:2015gda}.

If the black hole spin is misaligned with the orbit, then there are additional
effects of the spin~\cite{Foucart:2010eq,Foucart:2012vn,Kawaguchi:2015bwa}.
Spin-orbit coupling leads to orbital precession--the neutron
star and black hole ``bob'' above and below the initial orbital plane. 
For high misalignment angles ($\gtrsim 40^o$) and high black hole spin,
there can be notable
differences in the post-merger state.  If the neutron star disrupts,
the swath of matter is not restricted to one plane, so it can avoid hitting
itself, at least for a time, as it wraps around the black hole, delaying
the formation of a disk.  (See bottom panel on Fig~\ref{fig:merger}.)
Disks from significantly misaligned systems
are themselves misaligned, orbiting in a plane inclined with respect to
the equator defined by the black hole's spin, although this inclination
angle decreases on a timescale of tens of milliseconds.  Disk-black hole spin
misalignment can also lead to precession of the jet, which could produce
a periodicity in an observed GRB signal as the angle between direction of
jet and direction to observer oscillates.  Since binary neutron star mergers
do not seem to have a means of producing such misalignments, this
precession-induced GRB periodicity has been suggested as a way of
distinguishing GRBs from BHNS mergers~\cite{Stone:2012tr}.

In the outgoing tidal tail, some matter will be unbound and some bound.
The unbound matter contributes to outflows discussed below in
Section~\ref{sec:outflows}.  
The bound material eventually falls back onto the central black hole plus
disk.  Treating the bound outflow as following highly eccentric Keplerian
orbits, one can compute the fallback time, and from that the mass inflow
rate $\dot{M}_{\rm fallback}$.  This is found to follow a $t^{-5/3}$ power
law~\cite{Rosswog:2006rh,Chawla:2010sw,Brege:2018kii}.  This is explained
as follows~\cite{Rees:1988abc,Phinney:1989abc,Lodato:2008fr}.  The specific
orbital energy $e$ is related to the
semimajor axis $a$ and the period $t$ by $e\sim a^{-1}\sim t^{-2/3}$.  Then
the mass falling back in time interval $dt=dP$ is
$dM=(dM/de)(de/dt)dt\sim t^{-5/3}(dM/de)dt$.  If the ejecta/fallback mass
distribution is nonzero at $e=0$, then zooming into a narrow enough range
of $e$ near zero (corresponding to very late fallback times), the distribution
will look flat, and the fallback rate will asymptote to $\dot{M}\sim t^{-5/3}$.
In fact, $dM/de$ of bound ejecta is pretty flat, so most of the fallback
obeys the $t^{-5/3}$ rate~\cite{Kyutoku:2021icp}.

In the first
tens of milliseconds, fallback onto the disk can exceed accretion into
the black hole and significantly perturb and grow the disk.  The power law
falloff in $\dot{M}_{\rm fallback}$ would suggest that fallback can persist
till late times, later than the (viscous timescale) lifetime of the postmerger
disk, so that fallback has been considered as a source of late-time
($\sim$ hour) extended emission from some GRBs.  However,
two considerations alter the expectation of an uninterrupted power law
evolution of $\dot{M}_{\rm fallback}$.  First, the
accretion disk itself generates outgoing winds which would be expected
to interfere with fallback flows.  Indeed, a study by
Fern\'andez~{\it et al}~\cite{Fernandez:2014bra} on the interplay between
disk winds and dynamical ejecta shows that, for a $Q\approx 7$ merger,
disk winds can overcome and suppress fallback for times after about 100\,ms. 
Second, the assumption of Keplerian (ballistic) trajectories requires
that pressure forces in the bound ejecta remain small.  However, r-process
nucleosynthesis will, on timescales of order $\sim$second, heat the
fallback, providing thermal pressure to frustrate
fallback~\cite{Metzger:2010abc}.  This is not entirely bad news for an
extended emission model, though, since simulations of r-process heated
fallback find, for sufficiently massive post-merger black holes
($>6-8\,M_{\odot}$, probably true for most BHNS post-mergers), a temporary
pause in fallback
rather than a permanent stop, which is closer to observed GRBs followed
by extended emission than an uninterrupted power law~\cite{Desai:2018rbc}.
From these two considerations, fallback and associated late-time emission
may be complicated and $Q$-dependent.

Shocks at merger time allow the newborn disk to radiate neutrinos.  The
importance of weak interactions and neutrino emission depends on the
mass accretion rate.  If the accretion rate is above an ``ignition threshold''
of around $10^{-3}M_{\odot}\,\rm s^{-1}$, the disk will be efficently
neutrino cooled~\cite{De:2020jdt}.  If not, it will be advective.  
Eventually, even a disk with initial $\dot{M}$ will deplete sufficiently
to fall below this level, and neutrino emission loses its thermal importance.

Weak interactions and the ignition threshold also determine the composition
evolution.  Initially,
the disk matter is very neutron rich ($Y_e\approx 0.1$), like the neutron
star it came from.  Thus, $n+e^+$ reactions are more frequent than
$p+e^-$ reactions (cf. Eq~\ref{eq:epnnu} and~\ref{eq:enpnu}).
The fraction of nucleons that
are protons would then be expected to rise as electron antineutrino
emission dominates over electron neutrino emission.  This is, indeed,
what simulations find in the early ($\sim 10$\,ms) post-merger evolution.
After this, in efficiently neutrino cooled disks, $Y_e$ begins to
{\it decrease} again, settling back at around
0.1.  The reason is that the equatorial density becomes high enough
that the electron-positron gas becomes mildly degenerate:
$\mu_{e^-}/k_BT\approx 1$.  Equilibrium of
electron-positron pair creation/annihilation forces
$\mu_{e^+}=-\mu_{e^-}$.  The electron and positron distribution functions
differ only in the $\mu_{e^\pm}/kT$ term in the Fermi-Dirac
function, so as the electron-positron gas starts to become
degenerate, 
positrons become scarce compared to electrons, reducing the rate of
$n+e^+\rightarrow p+\overline{\nu}_e$~\cite{Beloborodov:2002af,Deaton:2013sla}.  The disk does not cool to become very degenerate
($\mu_{e^-}/k_BT\gg 1$) because strong degeneracy reduces the neutrino
luminosity, causing the
temperature to rise, so this process of approach to mild degeneracy with
$Y_e\approx 0.1$ is self-regulating~\cite{Beloborodov:2002af,Siegel:2017jug,De:2020jdt}.
Below the
ignition threshold, the degeneracy and regulation process does not
operate.  A bit later, as disk density and accretion rate decrease further,
weak interactions effectively shut off to the extent that $Y_e$ is
frozen in fluid elements (as long as NSE persists).  This extended
post-neutrino phase will turn out to be crucial for outflows (see
Section~\ref{sec:outflows} below).

\section{Magnetic field, jets, and gamma ray bursts}
\label{sec:MHD}

MHD simulations of BHNS post-merger evolution focus both on the magnetic
field's role in causing the turbulent ``viscosity'' and in the generation
of polar jets.  If the jet is generated by the Blandford-Znajek mechanism, the power
output $L_{\rm BZ}$ will be proportional to $\Phi^2$, where $\Phi$ is the
magnetic flux on the black hole horizon:
\begin{equation}
  \Phi = \int_H \left|B\cdot n\right| dA
\end{equation}
(Without the absolute value sign, field lines going in the horizon in
one hemisphere would cancel field lines coming out the other.)  Thus,
powerful jets require large magnetic flux on the horizon.  Accretion
from the disk can advect onto the horizon, and this flux can accumulate
with time.  The pressure of the surrounding disk is also needed to confine
the near-horizon field.  If the flux reaches
$\Phi=\Phi_{\rm MAD}\approx 50\dot{M}^{1/2}M_{\rm BH}$, the field will be
strong enough to oppose further accretion, leading to a magnetically arrested
disk (MAD) state~\cite{Narayan:2003by,Tchekhovskoy:2011abc}.  MAD accretion
releases energy very efficiently,
but since once $\Phi$ reaches $\Phi_{\rm MAD}$ it is subsequently stuck at
$\Phi\approx \Phi_{\rm MAD}$, the subsequent $L_{\rm BZ}$ is tied to the
(declining) mass accretion rate.

During and shortly after merger, several processes will amplify the
magnetic field inside the incipient disk.  First, the field will grow
due to {\it magnetic winding} (also called the ``$\omega$-effect'' in
dynamo literature):  as a simple consequence of the fact that
magnetic field lines are ``frozen into'' a conducting fluid, differential
rotation/shear will ``wind up'' field lines.  Winding will lead to a
linear in time growth of the toroidal component of the field ($B^{\phi}$).

The other sources of magnetic field growth are MHD instabilities.  For
example, as tidal disruption turns the star into a stream swirling around
the black hole, this wound spiral arm shears against itself,
triggering the Kelvin-Helmholtz
instability (KHI).  Vortices will form the shear interface in which magnetic
fields will quickly wind up~\cite{Kiuchi:2015qua,Izquierdo:2024rbb}.
%The growth rate of the KHI is proportional to
%$\lambda^{-1}$, meaning faster growth on smaller scales, making the KHI a
%serious challenge for both BHNS and NSNS simulations.

The most well-known is the {\it MRI}, which
is triggered in a rotating, conducting, weakly magnetized ($B^2\ll P$) fluid
whenever
\begin{equation}
  \varpi \partial_{\varpi}\Omega^2 + N^2 < 0\ ,
\end{equation}
where $\varpi$ is
the cylindrical radius (distance from the orbital axis),
$\Omega$ is the orbital angular frequency, and $N$ is the
Brunt-V{\"a}is{\"a}l{\"a} frequency~\cite{BH:1998}.  In a disk, the
first term will
dominate, and since $\Omega^2$ always decreases with $\varpi$ in disks, the
condition will be satisfied.  The MRI is a linear instability leading
to exponential growth at rate $\sim\Omega$ of the B-field.  The fastest
growing unstable mode has wavelength $\lambda_{\rm MRI}\sim v_A/\Omega$, where
$v_A=B/\sqrt{4\pi\rho}$ is the Alfven speed.  Thus, stronger fields (so long as they
are not strong enough to suppress the MRI) have larger $\lambda_{\rm MRI}$,
and numerical relativity simulations usually insert unphysically large
initial magnetic fields of order $10^{16}$\,G in order to be able to
numerically resolve the MRI in the subsequent merger.  This is justified
by pointing out that the inserted magnetar-level field is still dynamically
weak in the disk ($B^2/P\sim 10^{-2}$), in that there is much room for it to grow
to reach equipartition, so hopefully details of the seed field will not
imprint the field when it has saturated.

The MRI drives turbulence,
whereby energy injected at $\lambda_{\rm MRI}$ cascades to smaller scales,
and if one does not resolve a sufficient portion of the inertial range
at scales beneath $\lambda_{\rm MRI}$, turbulent transport effects might
be improperly modeled.  The highest resolution MHD BHNS simulations,
by Kiuchi~{\it et al}~\cite{Kiuchi:2015qua} and by
Izquierdo~{\it et al}~\cite{Izquierdo:2024rbb}
(minimum grid spacing 0.12\,km, the latter study including subgrid turbulence
modeling, both studies omitting neutrinos) fail to demonstrate numerical
convergence, with the former studying finding at higher resolution much
greater near-BH disk heating and early-time outflow rate and the latter
study much faster initial magnetic field growth driven by the KHI.

Early BHNS merger simulations with MHD took the
initial magnetic field to consist of poloidal loops with axis aligned with
the orbit and with field confined inside the
neutron star.  These simulations found that the field in the post-merger
disk quickly wound into a toroidally dominated configuration, with no
observable jets, at least on the tens of millisecond timescales
investigated~\cite{Chawla:2010sw,Etienne:2011ea,Etienne:2012tmf}.
(However, MRI-generated poloidal field can dominate in parts of the low
density region~\cite{Most:2021ytn}.)  Simulations of black hole accretion
disks have looked into the
differences when one begins with a poloidal vs. a toroidal seed field.
While the MRI inside the disk is found to be similar in both cases, jet
launching requires poloidal magnetic field reaching the black
hole~\cite{Beckwith:2007sr}.  Thus, it is not surprising that these
toroidal field-dominated merger outcomes did not quickly produce jets.

The University of Illinois Ubana-Champaign group has investigated the
effect of the geometry of the neutron star's initial magnetic field on
the post-merger field.  If the confined magnetic field is
tilted with respect to the orbital axis, there can be a somewhat larger
early post-merger poloidal fields, but still insufficient to generate an
early post-merger jet~\cite{Etienne:2012tmf}.  On the other hand, jets
can more quickly ($\approx$100\,ms) emerge from a BHNS
merger involving a neutron star with pre-merger poloidal field extending
outside the neutron star, because in such a case more magnetic field survives
the merger and can be wound above the poles into an incipient
jet~\cite{Paschalidis:2014qra,Ruiz:2018wah}.  These were, in fact, the
first BHNS simulations to successfully demonstrate jet formation.

\begin{figure}
  \includegraphics[width=10cm]{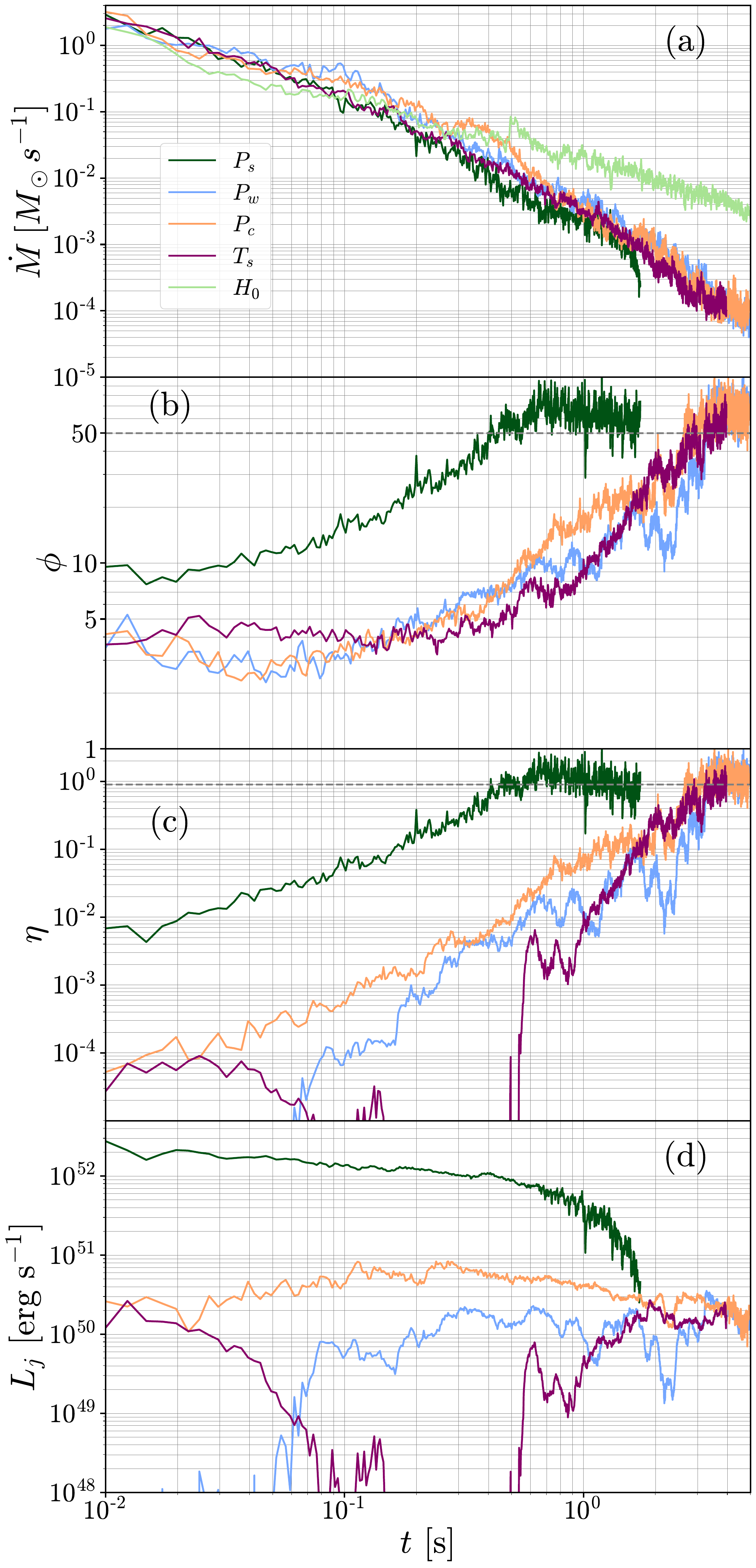}
  \caption
      {
        Electromagnetic energy extraction (measured at $r=5M_{\rm BH}$)
        for a $Q=2$,
        $\chi_{\rm BH}=0.6$ system with different choices of magnetic field
        inserted post-merger.  Figure from
        Gottlieb~{\it et al}~\cite{Gottlieb:2023est} under the terms of the
        Creative Commons Attribution 4.0 International License.
        {\bf Panel (a)}: Mass accretion rate
        {\bf Panel (b)}: Dimensionless magnetic flux
        $\phi = \Phi/\sqrt{\dot{M}M_{\rm BH}^2} $, where $ \Phi $ is the magnetic
        flux. Ultimately, all models turn MAD when $ \phi \approx 50$
        (dashed line).
        {\bf Panel (c)}: The jet launching efficiency
        {\bf Panel (d)}: The jet luminosity, $L_j = \eta\dot{M}$
      }
  \label{fig:MAD}
\end{figure}

Jets are still possible for disks with
toroidal initial fields, but there is the delay that the disk must first
generate a poloidal field by {\it dynamo} action, the process whereby
some of the toroidal field can be converted through small-scale turbulence
into a large-scale poloidal field~\cite{Moffatt:1978abc}.  Magnetic winding feeds poloidal
field into toroidal field, so the dynamo can be a self-sustaining cycle.
Distinctive oscillatory behavior is seen in the large-scale dynamo-driven
field.  In
black hole accretion systems, one can see a cyclic drift in the latitude of
toroidal field strength reminiscent of the ``butterfly diagrams'' in
solar sunspots~\cite{Hogg:2018zon,Hayashi:2021oxy}.  Disk simulations
with toroidal initial field find that dynamo poloidal field growth
can indeed happen~\cite{Liska:2018btr,Christie:2019lim}, although the
jets that eventually form might be more intermittant (``striped'') than
those from disks with poloidal seed field~\cite{Christie:2019lim}.

A few very recent simulations have followed magnetized BHNS disks for
second timescales:  the work of Hayashi~{\it et al}~\cite{Hayashi:2021oxy,
  Hayashi:2022cdq} using the {\tt SACRA} code and the work of
Gottlieb~{\it et al}~\cite{Gottlieb:2023est,Gottlieb:2023sja} using the
{\tt H-AMR} code with initial data from hydrodynamic BHNS mergers with the
{\tt SpEC} code.  The two studies differed in several ways.  Notably,
the {\tt SACRA} simulations included neutrino effects and recombination
while the {\tt H-AMR} simulations did not, but the {\tt H-AMR} simulations
evolved for longer times. 
Here, we will present the main points of the picture emerging
from extended post-merger MHD simulations, encouraging readers to
consult the original papers for details.  Readers should note that some
papers only insert a seed magnetic field after
merger (e.g.~\cite{Nouri:2017fvh,Gottlieb:2023est,Gottlieb:2023sja}).
Magnetized BHNS mergers find that the field in the initial post-merger disk
is predominantly toroidal, so the early evolution of disks with poloidal
seed field should not be trusted much.  Also, electromagnetic luminosity
can be reported differently, e.g. Poynting luminosity
across the whole horizon, a measure of electromagnetic energy extraction
from the black hole, vs. isotropic equivalent luminosity to the Poynting
flux near the axis, a measure of jet strength relevant to an on-axis observer.

Dynamo activity is clearly seen in extended MHD simulations of BHNS
disks, both in the growth of poloidal field and in the distinctive
``butterfly'' effect drift in the toroidal field~\cite{Hayashi:2021oxy,
  Hayashi:2022cdq}).  The polar region after merger is not empty but
contains matter ejected during tidal disruption and merger as well as
early-time disk outflows.  This delays the formation of a magnetosphere
until the density inside a funnel region falls to
$\rho < b^2/8\pi$~\cite{Most:2021ytn,Hayashi:2021oxy}, and the incipient
jet must push through a cocoon of this matter~\cite{Gottlieb:2023est}.

The horizon magnetic flux $\Phi$, and hence the Poynting luminosity for the
Blandford-Znajek effect, accumulates
early and is fairly constant for order seconds, excepting fluctuations due
to turbulence
and dynamo cycles.  During this time, the disk depletes, and the accretion
rate falls off like $\dot{M}\propto t^{-2}$.  Thus the MAD criterion ratio
$\Phi/(\dot{M}M_{\rm BH})$ grows.  Eventually, it is inevitable that the
disk will enter a MAD state.  The nominal efficiency of accretion
($L_{\rm BZ}/\dot{M}$) grows (``output'' holding steady while ``input''
decreases), reaching order unity as the disk reaches MAD state.  As
$\dot{M}$ drops, gas pressure at the edge of the funnel region decreases,
so the gas loses ability to confine the jet.  Even before the MAD
criterion is satisfied, this loss of confining pressure causes the jet
to broaden, leading to a corresponding decrease of flux observed by an
observer in the jet's path~\cite{Hayashi:2021oxy}.  Once the disk becomes
MAD, the jet luminosity is constrained to decrease along with $\dot{M}$
at a rate $\propto t^{-2}$, which we should be able to observe in short
GRBs.  These features of the evolution are shown in Figure~\ref{fig:MAD}.

If it is true that the ultimate falloff of GRB jet luminosity is caused
by the growing relative strength of magnetic forces on a depleting disk, then
there is a connection between the post-merger disk mass (which determines
$\dot{M}$), the jet luminosity, and the lifetime after which the jet luminosity
drops, as illustrated in Figure~\ref{fig:MAD}.  For a given disk mass (and hence
$\dot{M}(t)$), a more luminous jet would reach this time at which decay
commences sooner.  Considering only the MAD criterion, this is when the
$L_{\rm BZ}$ and $\dot{M}$ curves cross~\cite{Gottlieb:2023sja}.

Magnetic fields are not the only mechanism that has been considered for
launching jets.  Another energy source comes from the annihilation of
neutrinos and antineutrinos in the (neutrino) optically-thin polar region. 
The luminosity from neutrino annihilation for a hot, high $\dot{M}$ (but
neutrino transparent) disk is roughly
$L_{\nu\overline{\nu}}\sim 10^{52} (\dot{M}/M_{\odot}s^{-1})^{2.25}$ erg\,s${}^{-1}$~\cite{Zalamea:2011abc}.  For high accretion rates, of order a percent of
the neutrino luminosity can be deposited in this way, but naturally, this
source of energy shuts off when
the accretion rate drops below the weak interaction ignition accretion rate,
and is probably negligible for
$\dot{M}<10^{-2}M_{\odot}\,s^{-1}$~\cite{Popham:1998ab}.  This is one problem
for the $\nu\overline{\nu}$ GRB model.  Another obstacle to relativistic
jets is the inertia of matter ejected during merger polluting the polar
regions (``baryon loading'').  Radiative hydrodynamic
evolutions suggest that this ejecta will prevent $\nu\overline{\nu}$
annihilation from accelerating polar outflows to relativistic speeds
in NSNS remnants~\cite{Just:2015dba}.  The same paper concludes that
BHNS disks (which have somewhat cleaner polar regions after merger)
can produce GRB fireballs, but the energies and durations are
too low to explain most short GRBs.  It remains possible that the
neutrino mechanism is important early on.  Since, according to high resolution
MHD BHNS simulations, magnetic jets take at least tens of milliseconds to
form~\cite{Kiuchi:2015qua}, the character of the relativistic outflow
might change from fireball to Poynting flux dominated
(cf.~\cite{Barkov:2011jf}).  It should also be remembered that, even if
$\nu\overline{\nu}$ annihilation cannot drive a relativistic outflow,
it might contribute to mildly relativistic winds
(cf.~\cite{Fujibayashi:2017xsz}), a topic to which we now turn.

\section{Outflows, Kilonovae, and Radio Flares}
\label{sec:outflows}

Matter from the neutron star that ends up being ejected will decompress
and evolve into some kind of ``normal'' matter; this evolution turns out
to be of great interest for electromagnetic signals and production of
heavy elements.  When $\rho$ and $T$ are low enough that outgoing gas
leaves NSE, many nucleons will have collected into heavy nuclei.  However,
if $Y_e$ is low (many more neutrons than protons), there will be a large
number of ``extra'' free neutrons as well.  In this case, nuclei will
rapidly capture neutrons, bringing them close to the neutron drip line,
i.e. the most neutrons a nucleus can hold for given number of protons--on a
space of isotopes labeled by $(A,Z)$, the maximum of $A$ for given $Z$.
These nuclei can eventually form the ``r-process'' elements.  The
distribution of nuclei produced by the r-process has three peaks at
$A\sim$70-80, 120-130, and 190-200.  Notably, it produces lanthanides
and actinides (located past the second peak).  Absent rapid neutron
capture (i.e. for higher $Y_e$), nuclei stay closer to the valley of
stable isotopes (a swath in the $A$-$Z$ plane); there can be s-process
nucleosynthesis in this regime.  For high-$Y_e$ outflow, the r-process
does not proceed far enough to produce significant amounts of
elements beyond the second or even first peak; in particular lanthanides
and actinides are not efficiently produced.

The possibility that tidal disruption in BHNS mergers could produce a
large mass of unbound outflow which might then undergo
r-process nucleosynthesis was pointed out by Lattimer and Schramm
in 1976~\cite{Lattimer:1976abc}.  Li and Paczynski noticed that radioactive
decays in ejecta could power a bright electromagnetic
transient~\cite{Li:1998bw}.  However, it was only sometime afterward that
models were created with realistic heating rates and
opacities~\cite{Metzger:2010abc,Barnes:2013wka,Kasen:2013xka,Kasen:2017sxr}.
Opacity turns out to be crucial--a more opaque gas will trap photons longer,
so that they escape on a longer timescale and at lower energy.  Lanthanides,
if present, introduce a thicket of absorption lines due to their
partially-filled valence f shells, greatly increasing the bound-bound
opacity.  A lanthanide-rich ejecta glows in the near infrared on timescale of
a week (``red kilonova''); a lanthanide-poor ejecta glows in the optical-UV on
day timescale (``blue kilonova'').  The transition between one and the other
is at about $Y_e\approx 0.25$~\cite{Wanajo:2014abc,Lippuner:2015gwa,
  Tanaka:2019iqp}.

As the outflow spreads and decompresses further, it interacts with the
background interstellar medium (ISM); electrons are accelerated in the shock
moving into the ISM and emit synchrotron radiation.  This results in a
distinct signal, a radio flare lasting years to decades~\cite{Nakar:2011cw,
  Hotokezaka:2015eja,Hotokezaka:2016clu}.  This long timescale is set by
the deceleration time, the time it takes the ejecta to sweep up roughly
its own mass in ISM.  The timescale and
strength of these flares depends both on the kinetic energy of the ejecta
and the density of the surrounding ISM.  Thus, the properties of outflows
determine multiple signals.

Numerical simulations predict the mass of outflow and its distribution in
speed, entropy, and $Y_e$.  One issue is how to identify
unbound matter.  In a stationary spacetime, $u_t$ is constant on geodesics,
so a $u_t<-1$ geodesic can be identified as unbound.  Furthermore, $-u_t$
can be identified as the asymptotic Lorentz factor.  However, the spacetime
is not stationary in the early merger, and--a more serious problem--the
outflow is not geodesic, at least within the grid and timeframe which most
simulations can afford to maintain, because of the persistence of pressure
forces (e.g.~\cite{Haddadi:2022qcu}).  In a stationary flow (not necessarily
geodesic), the Bernoulli
parameter $hu_t$ is constant on fluid elements.  At asymptotically large
distances and late times, $\rho_0\rightarrow 0$, $h\rightarrow h_{\infty}$,
so for such a flow, one can tag $(h/h_{\infty})u_t<-1$
fluid elements as unbound and $-(h/h_{\infty})u_t$ as the asymptotic Lorentz
factor, but, unfortunately, the outflow is not stationary either.  One must
also worry about the effect of r-process heating, currently not accounted
for in most merger and post-merger simulations.  For more on the subtleties of
identifying unbound matter and its asymptotic speed, see
Foucart~{\it et al}~\cite{Foucart:2021ikp}.

\begin{figure}
  \includegraphics[width=10cm]{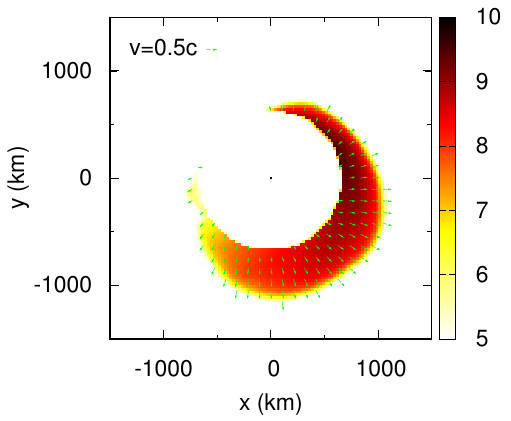} \\
  \includegraphics[width=10cm]{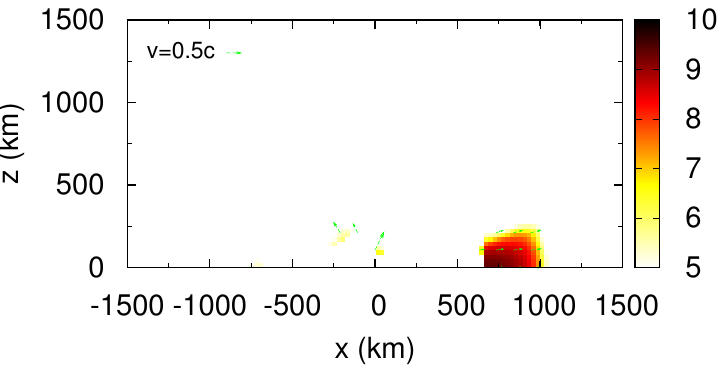}
  \caption
      {
        Concentration of dynamical ejecta for a $Q=5$, $\chi_{\rm BH}=0.75$,
        aligned spin BHNS merger.  {\bf Top:} 
        equatorial plane. {\bf Bottom:} a meridional plane.  Color
        indicates density.  Reprinted figure with permission
        from Kyutoku, Ioka, and Shibata, Phys. Rev. D {\bf 88},
        041503 (2013)~\cite{Kyutoku:2013wxa}.  Copyright 2013 by the
        American Physical Society.
      }
  \label{fig:ejecta}
\end{figure}

Newtonian and relativistic simulation confirm that
BHNS mergers produce large ejecta masses, up to
$\sim 10^{-1}M_{\odot}$~\cite{Rosswog:2005su,Kyutoku:2013wxa,Foucart:2012vn,
  Kyutoku:2015gda,Lovelace:2013vma,Kyutoku:2017voj,Brege:2018kii}.
Dynamical ejecta
masses can thus be much larger than in NSNS systems, mainly because of
the higher mass ratios $Q$ for BHNS binaries, rather than because of
something distinctive about the presence of a black hole.  (Asymmetric
NSNS systems also tend to produce more dynamical ejecta than equal mass
systems.)  Since the ejecta is produced by tidal disruption of the neutron star
before merger, the ejecta is cold and little affected by neutrino
(weak interaction) processes~\cite{Kyutoku:2017voj,Roberts:2016igt},
so its electron fraction is essentially what it was in the pre-merger
neutron star, about $Y_e\approx 0.05$.  Such extremely neutron-rich
material will produce 2nd and 3rd peak r-process elements and produce a
red kilonova.  Typical asymptotic velocities are $\sim 0.2$--$0.3c$.
BHNS dynamical ejecta is extremely anisotropic, much more
so than NSNS ejecta, with outflow concentrated within $20^\circ$ of the
equatorial plane and sweeping out only about half of this plane ($180^\circ$). 
(See Fig.~\ref{fig:ejecta}.) 
The ejecta thus has an overall linear momentum, so the black hole remnant
can get a kick of $\sim 100\,\rm km\ s^{-1}$, a larger effect than
the gravitational wave recoil~\cite{Kyutoku:2013wxa,Foucart:2012vn,
  Kyutoku:2015gda}.  Further consequences of this overall direction
of ejecta motion will be discussed below.  It should be pointed out,
though, that BHNS systems with significantly misaligned black hole spin
can produce dynamical ejecta extending over $360^\circ$~\cite{Kawaguchi:2015bwa}.

An additional component of outflow comes later from the accretion disk,
especially after neutrino cooling has diminished in importance, and the
disk has become advective.  These disk wind outflows have been studied
in Newtonian physics in 2D, with the black hole modeled as a central
potential and MHD turbulence modeled as a shear viscosity, in a series
of papers by Fernandez and collaborators~\cite{Fernandez:2013,Fernandez:2014cna,Fernandez:2014bra,Fernandez:2016sbf,Fernandez:2020oow}.  There have also been
a small number of 2D relativistic viscous hydrodynamic studies, which
find qualitatively similar results~\cite{Fujibayashi:2020jfr,Fujibayashi:2020,Haddadi:2022qcu}.  Outflow gas is unbound from the disk
by thermal pressure, so the key to understanding disk winds is to understand
how thermal energy is added to the disk.  Viscosity provides the major source
of heating; since viscosity is considered a model for turbulence, this viscous
heating can be interpreted as dissipation of turbulent energy on small scales.
Heat generated in the interior of the disk can trigger convection which
transports this heat to the surface.  (Thus, even simulations which don't
explicitly include magnetorotational turbulence often show disk snapshots which
look quite turbulent.) 
Thermal energy is also boosted by recombination of nucleons into nuclei, with
the accompanying release of nuclear binding energy.  (Recall, this is not
technically ``heating'' since it is reversible and adiabatic so long as the
fluid remains in NSE.  It is automatically
included in simulations that use temperature-dependent equations of state
and use a conservative formulation of the fluid or MHD equations.)

A significant fraction of the disk's mass can be ejected in these winds, with
10\% to 20\% of the initial disk mass being typical.  An exploration of the
parameter space of disks~\cite{Fernandez:2020oow} shows that the fraction
of disk mass ejected decreases with the compactness of the initial disk
(proportional to the black hole mass divided by the radius of the initial
density peak).  The dependence on disk compaction can be understood by
remembering that most of the outflow launching happens after weak interaction
freeze-out, when accretion becomes advective.  For a more compact disk, more
of the mass will accrete before this time, meaning that for a less compact
disk, more mass will be remaining when strong outflows commence.
Ejected mass fraction also decreases with the initial disk mass.
Ejecta from disk outflows is found to be significantly slower than dynamical
ejecta, with $v\approx 0.05c$ being a typical measurement.  Unlike the
dynamical ejecta, this ejecta undergoes significant neutrino processing
from its time in the disk, and average $Y_e\approx 0.3$.  Of course, within
the disk outflow from a single merger, there will be gas with a range of
$Y_e$, but more than half will be lanthanide and actinide-poor and thus
lower opacity.

Viscous hydrodynamics is an approximation meant to model MHD turbulence.
How well outflows from viscous evolution match those of full MHD evolution
has been tested by comparing results of 3D MHD simulations to those of 2D
viscous hydrodynamics, with both simulation types including neutrino cooling
and nuclear recombination.  One finds that MHD simulations
produce late-time thermal winds that are very similar in mass, velocity,
and composition to the outflows in 2D viscous hydrodynamics simulations
for alpha viscosity in the range $\alpha_{\rm SS}\approx 0.03$--0.1.
Simulations with poloidal initial field (with ratio of gas pressure
to magnetic pressure of $\beta\sim 10^2$) find an additional,
early-time
magnetically-driven ejecta, amounting to another 20\% of the disk's mass,
with higher velocity and lower $Y_e$ than the later
wind~\cite{Fernandez:2018kax}.  This early-time outflow is, however,
sensitive to the choice of seed field~\cite{Christie:2019lim}.  It is
essentially removed if the initial poloidal field is weak ($\beta\sim 10^3$)
or if the initial field is toroidal.  BHNS merger MHD simulations indicate the
early post-merger field is in fact predominantly toroidal, and one does
not observe the early outflow in these simulations~\cite{Hayashi:2021oxy}.
(However, outflow might appear only at sufficiently high
resolution~\cite{Kiuchi:2015qua}, so one should be cautious with these
conclusions.)

We have seen that disrupting BHNS mergers will, except for systems
with low-mass black holes, have as much or more dynamical ejecta than
disk ejecta.  Furthermore, we have seen that the dynamical ejecta can be
highly asymmetric--often concentrated near the orbital equator and to one
side.  Both the mass and asymmetry of dynamical ejecta distinguish typical
BHNS from typical NSNS outflows.  They have observable
consequences.

Consider the asymmetry of ejecta.  The flatness of the ejecta means
photons are able to random walk to the vertical surface faster than they
would be able to escape a spherical distribution.  This results in a
shorter time to peak emission, higher ejecta temperature at this time,
higher peak luminosity, and bluer emission~\cite{Kyutoku:2013wxa,
  Tanaka:2013ixa}.  It has even been suggested that the difference in color
for a given luminosity might provide a way to distinguish BHNS from NSNS
mergers from their kilonovae~\cite{Tanaka:2013ixa}.  The asymmetry of
ejecta motion, on the other hand, causes Doppler effects, so that the
brightness and color depends on the viewing angle--whether the ejecta
is moving toward or away from the observer~\cite{Darbha:2021abc}.

The overall kilonova emission will depend on emission from and opacity of
both dynamical and disk outflow components and their spatial arrangement.
The dynamical ejecta will be farther out from the black hole and equatorially
concentrated, while the disk outflow will be more isotropic.  Thus,
there may be a low-opacity disk outflow producing a blue
emission, but it might only be visible from certain viewing angles, because
in equatorial directions, the blue emission is blocked by the more-opaque
dynamical ejecta between the disk outflow and the viewer.  Some work has
been done in figuring out
how the different ejecta component effects combine at different viewing
angles to produce overall light curves~\cite{Fernandez:2014bra,
  Fernandez:2016sbf,Darbha:2020lhz,Kawaguchi:2019nju}.

The distinct features of BHNS dynamical ejecta also affect the later radio
flare.  A flat outflow takes longer than a spherical outflow of the same
mass to sweep up its own mass, so the peak time might be delayed.  Also,
the overall motion of the ejecta in one direction results in a proper motion
of the radio source which could possibly be detected~\cite{Kyutoku:2013wxa}.

\section{Conclusions}
\label{sec:conclusions}

BHNS mergers involve physics similar to that of NSNS mergers and have the
potential to produce similar types of signals.  They are distinguished from
NSNS mergers in that higher mass ratio $Q$ is expected, the post-merger
central remnant is definitely a black hole, the black hole is likely more
massive than that produced by NSNS post-merger collapse, greater dynamical
ejecta masses are possible, this dynamical ejecta is colder, more neutron-rich,
and more anisotropic, and the post-merger polar region is likely to
be less baryon loaded.  These differences may have observational consequences.
The challenge of MHD simulations comes from the range of time and length
scales that must be covered.  However, general aspects of the merger and
post-merger evolution are coming to be known with confidence, because they
follow from general principles:  the dependence of tidal disruption on
binary parameters, the early importance but eventual shut-off of
neutrino-emitting weak interactions, and the natural evolution of the
disk toward a magnetically arrested state.

Nevertheless, no current simulation of the post-merger evolution should
be considered definitive, because none contains all of the ingredients
needed to be fully trustworthy.  MHD and neutrino emission are both
crucial to the post-merger emission, and while many studies include
one or the other, few include them both.  Those of which the author is
aware include~\cite{Nouri:2017fvh,Most:2021ytn,Hayashi:2021oxy,
  Hayashi:2022cdq}.  The first two of these used neutrino leakage, which
is only quantitatively reliable for optically thin disks.  However, even
including both magnetic fields and neutrino transport is insufficient. 
First, there are the difficulties of resolving turbulence induced by the
MRI.  Convergence studies without neutrino effects (which are thus
{\it physically} unreliable) indicate that grid spacing of $\Delta x<0.2$\,km
is needed to capture early outflows~\cite{Kiuchi:2015qua}.  The studies cited
above that include neutrino effects do not see these outflows or this
resolution sensitivity, but they also have not gone to such high
resolution.  Furthermore, the neutrino treatment in these studies did not
include neutrino-antineutrino annihilation and flavor oscillations.
Simulations continuing for seconds and tracking ejecta to large distances
face the more important problem of the breakdown of nuclear statistical
equilibrium, which (because of r-process heating) cannot adequately be
relegated to post-processing.

In addition to these modeling issues, future simulations must more adequately
explore the BHNS parameter space.  Much of the work reviewed in this chapter
has concentrated on low-to-moderate mass ratio systems with ``canonical''
neutron star masses of 1.3--1.4$M_{\odot}$ and aligned black
hole spin.  Properly calibrating BHNS gravitational wave models will require
more adequately sampling a fuller range of the possible masses and spins.
We should be sure, for example, that the conclusions we have drawn regarding
characteristically asymmetric BHNS ejecta are not too sensitive to an
assumption that the black hole spin misalignment angle is small.  We should
be careful in extrapolating the conclusions of our few seconds-long MHD
simulations~\cite{Hayashi:2022cdq,Gottlieb:2023est} carried out from two
sets of masses and spins to the whole range of disrupting BHNS systems.
As daunting as the task may be, we must carry out seconds-long radiation
MHD simulations covering a wide range of BHNS systems.

\begin{acknowledgement}
  I am grateful to Francois Foucart, Ore Gottlieb, Koutarou Kyutoku,
  Nishad Muhammed, Stuart Shapiro, and Samuel Tootle for viewing a draft and providing useful comments. I also
  thank the authors of references~\cite{Kyutoku:2011vz},
  \cite{Paschalidis:2013jsa}, and \cite{Kyutoku:2013wxa}
  for permission to reuse figures~\ref{fig:bhnswaves},
  \ref{fig:magnetosphere}, and~\ref{fig:ejecta}.  I am grateful for
  support from the NSF through grant PHY-2110287 and from NASA through
  grant 80NSSC22K0719.
\end{acknowledgement}

\bibliographystyle{sn-aps}
\bibliography{References}

\end{document}